\documentclass[11pt]{article}
\usepackage{amsmath}
\usepackage{amsfonts}
\usepackage{amssymb}
\usepackage{graphicx}
 \usepackage{epstopdf} 

\ifx\pdfoutput\@undefined\usepackage[usenames,dvips]{xcolor}
\else\usepackage[usenames,dvipsnames]{xcolor}
\IfFileExists{pdfcolmk.sty}{\usepackage{pdfcolmk}}{} 
\fi
\usepackage[plainpages=false,pdfpagelabels,pagebackref=false,naturalnames=true,hyperindex=true,pdftitle={What is Computation?},pdfauthor={Hector Zenil}]{hyperref}
\hypersetup{colorlinks=true,
urlcolor=Cerulean,linkcolor=BrickRed,citecolor=RoyalBlue,a4paper,
  pdfpagemode=None,
  pdfstartview=FitH}
\usepackage[all]{hypcap}

\begin{document}

%\def\bibsection{\section*{References}}

%\usepackage{epsf}

% first colour for latex or pdflatex
%\ifx\pdfoutput\@undefined\usepackage[usenames,dvips]{xcolor}
%\else\usepackage[usenames,dvipsnames]{xcolor}
% and fix pdf colour problems
%\IfFileExists{pdfcolmk.sty}{\usepackage{pdfcolmk}}{}
%\fi
%\usepackage[plainpages=false,pdfpagelabels,pagebackref=false,naturalnames=true,hyperindex=true,pdftitle={Two-Dimensional Kolmogorov Complexity and an Empirical Validation of the Coding Theorem Method by Compressibility},pdfauthor={Hector Zenil}]{hyperref}
%\hypersetup{colorlinks=true,
%urlcolor=Cerulean,linkcolor=BrickRed,citecolor=RoyalBlue,a4paper,
% pdfpagemode=None,
% pdfstartview=FitH}
%\usepackage[all]{hypcap}

%\begin{frontmatter}

\title{\vspace{-4cm}Two-Dimensional Kolmogorov Complexity and an Empirical Validation of the Coding Theorem Method by Compressibility\thanks{Corresponding author: \url{hectorz@labores.eu}.}}
\author{Hector Zenil$^{1,2,6}$, Fernando Soler-Toscano$^{3,6}$, Jean-Paul Delahaye$^{4,6}$\\and Nicolas Gauvrit$^{5,6}$\\
$^1$ Unit of Computational Medicine, Department of Medicine Solna,\\SciLifeLab, (Science for Life Laboratory), Centre for Molecular\\ Medicine and Karolinska Institute, Stockholm, Sweden.\\
$^2$ Department of Computer Science, University of Oxford, UK.\\
$^3$ Grupo de L\'ogica, Lenguaje e Informaci\'on,\\Universidad de Sevilla, Spain.\\
$^4$ CRISTAL (Centre de recherche en informatique, signal et \\automatique de  Lille), France.\\
$^5$ CHArt Lab, \'Ecole Pratique des Hautes Etudes, Paris, France.\\
$^6$ Algorithmic Nature Group, LABoRES, Paris, France.}

\date{}

\maketitle
                                             
%\title{Two-Dimensional Kolmogorov Complexity and Validation of the Coding Theorem Method by Compressibility\tnoteref{t1}}
% \tnotetext[t1]{Support material can be found at the \emph{Algorithmic Nature Group} website at \url{http://www.algorithmicnature.org}.}
%\author{Fernando Soler-Toscano$^1$, Hector Zenil$^2$\cortext[cor1]{Corresponding author: h.zenil@sheffield.ac.uk}, Jean-Paul Delahaye$^3$\\and Nicolas Gauvrit$^4$}

%\author[hector]{Hector Zenil\corref{cor1}}
%\author[fer]{Fernando Soler-Toscano}
%\author[jp]{Jean-Paul Delahaye}
%\author[nico]{Nicolas Gauvrit}

%\cortext[cor1]{Corresponding author: \url{hectorz@labores.eu}.}

%\address[hector]{Department of Computer Science, University of Sheffield, UK.}
%\address[fer]{Grupo de L\'ogica, Lenguaje e Informaci\'on, Universidad de Sevilla, Spain.}
%\address[jp]{Laboratoire d'Informatique Fondamentale de Lille, Universit\'e de Lille I, France.}
%\address[nico]{LDAR, Universit\'e de Paris 7, France.}

\begin{abstract}
We propose a measure based upon the fundamental theoretical concept in algorithmic information theory that provides a natural approach to the problem of evaluating $n$-dimensional complexity by using an $n$-dimensional deterministic Turing machine. The technique is interesting because it provides a natural algorithmic process for symmetry breaking generating complex $n$-dimensional structures from perfectly symmetric and fully deterministic computational rules producing a distribution of patterns as described by algorithmic probability. Algorithmic probability also elegantly connects the frequency of occurrence of a pattern with its algorithmic complexity, hence effectively providing estimations to the complexity of the generated patterns. Experiments to validate estimations of algorithmic complexity based on these concepts are presented, showing that the measure is stable in the face of some changes in computational formalism and that results are in agreement with the results obtained using lossless compression algorithms when both methods overlap in their range of applicability. We then use the output frequency of the set of 2-dimensional Turing machines to classify the algorithmic complexity of the space-time evolutions of Elementary Cellular Automata.\\

\end{abstract}

\section{Introduction}

The question of natural measures of complexity for objects other than strings and sequences, in particular suited for 2-dimensional objects, is an open important problem in complexity science and with potential applications to molecule folding, cell distribution, artificial life and robotics. Here we provide a measure based upon the fundamental theoretical concept that provides a natural approach to the problem of evaluating $n$-dimensional algorithmic complexity by using an $n$-dimensional deterministic Turing machine, popularized under the term of \textit{Turmites} for $n=2$, from which the so-called \textit{Langton's ant} is an example of a Turing universal \textit{Turmite}. A series of experiments to validate estimations of Kolmogorov complexity based on these concepts is presented, showing that the measure is stable in the face of some changes in computational formalism and that results are in agreement with the results obtained using lossless compression algorithms when both methods overlap in their range of applicability. We also present a \textit{divide and conquer} algorithm that we call \textit{Block Decomposition Method} (BDM) application to classification of images and space-time evolutions of discrete systems, providing evidence of the soundness of the method as a complementary alternative to compression algorithms for the evaluation of algorithmic complexity. We provide exact numerical approximations of Kolmogorov complexity of square image patches of size 3 and more, with the BDM allowing scalability to larger 2-dimensional arrays and even greater dimensions.

The challenge of finding and defining 2-dimensional complexity measures has been identified as an open problem of foundational character in complexity science~\cite{feldman1,shalizi}. Indeed, for example, humans understand 2-dimensional patterns in a way that seems fundamentally different than 1-dimensional~\cite{feldman}. These measures are important because current 1-dimensional measures may not be suitable to 2-dimensional patterns for tasks such as quantitatively measuring the spatial structure of self-organizing systems. On the one hand, the application of Shannon's Entropy and Kolmogorov complexity has traditionally been designed for strings and sequences. However, $n$-dimensional objects may have structure only distinguishable in their natural dimension and not in lower dimensions. This is indeed a question related to the lost in dimension reductionality~\cite{zenilkianitegner}. A few measures of 2-dimensional complexity have been proposed before building upon Shannon's entropy and block entropy~\cite{feldman1,andrienko}, mutual information and minimal sufficient statistics~\cite{shalizi} and in the context of anatomical brain MRI analysis~\cite{young,young2}. A more recent application, also in the medical context related to a measure of consciousness, was proposed using lossless compressibility for EGG brain image analysis was proposed in~\cite{casali}.

On the other hand, For Kolmogorov complexity, the common approach to evaluating the algorithmic complexity of a string has been by using lossless compression algorithms because the length of lossless compression is an upper bound of Kolmogorov complexity. Short strings, however, are difficult to compress in practice, and the theory does not provide a satisfactory solution to the problem of the instability of the measure for short strings. 

Here we use so-called \textit{Turmites} (2-dimensional Turing machines) to estimate the Kolmogorov complexity of images, in particular space-time diagrams of cellular automata, using Levin's Coding theorem from algorithmic probability theory. We study the problem of the rate of convergence by comparing approximations to a universal distribution using different (and larger) sets of small Turing machines and comparing the results to that of lossless compression algorithms carefully devising tests at the intersection of the application of compression and algorithmic probability. We found that strings which are more random according to algorithmic probability also turn out to be less compressible, while less random strings are clearly more compressible.

%The chief advantage of lossless compression algorithms is that they are a sufficient test of non-randomness. However, for short strings, which are usually the ones useful in practical applications, adding the decompression instructions to the compressed version makes the compressed string often, if not always, longer than the string itself, simply because the decompression instructions are at least equal in length to the original object. 

Compression algorithms have proven to be signally applicable in several domains (see e.g. \cite{li}), yielding surprising results as a method for approximating Kolmogorov complexity. Hence their success is in part a matter of their usefulness. Here we show that an alternative (and complementary) method yields compatible results with the results of lossless compression. For this we devised an artful technique by grouping strings that our method indicated had the same program-size complexity, in order to construct files of concatenated strings of the same complexity (while avoiding repetition, which could easily be exploited by compression). Then a lossless general compression algorithm was used to compress the files and ascertain whether the files that were more compressed were the ones created with highly complex strings according to our method. Similarly, files with low Kolmogorov complexity were tested to determine whether they were better compressed. This was indeed the case, and we report these results in Section~\ref{comparison}. In Subsection~\ref{eca} we also show that the Coding theorem method yields a very similar classification of the space-time diagrams of Elementary Cellular Automata, despite the disadvantage of having used a limited sample of a \emph{Universal Distribution}. In all cases the statistical evidence is strong enough to suggest that the Coding theorem method is sound and capable of producing satisfactory results. The Coding theorem method also represents the only currently available method for dealing with very short strings and in a sense is an expensive but powerful ``microscope" for capturing the information content of very small objects.

\section{Kolmogorov-Chaitin complexity}
\label{kolmo}

Central to algorithmic information theory (AIT) is the definition of algorithmic (Kolmogorov-Chaitin or program-size) complexity \cite{kolmo,chaitin}:

\begin{equation}
K_T(s) = \min \{|p|, T(p)=s\}
\end{equation}

 That is, the length of the shortest program $p$ that outputs the string $s$ running on a universal Turing machine $T$. A classic example is a string composed of an alternation of bits, such as $(01)^n$, which can be described as ``$n$ repetitions of 01". This repetitive string can grow fast while its description will only grow by about $\log_2(n)$. On the other hand, a random-looking string such as $011001011010110101$ may not have a much shorter description than itself.
 
%The measure was first conceived to define randomness and is today the accepted objective and universal mathematical measure of randomness, among other reasons because it has been proven to be mathematically robust (by virtue of the fact that several independent definitions converge in it), although it is worth noting that there are important differences between infinite and finite randomness that we will not discuss here (among other reasons because we are only concerned with finite randomness in this paper).

 \subsection{Uncomputability and instability of $K$}
 \label{compress}
 
 A technical inconvenience of $K$ as a function taking $s$ to the length of the shortest program that produces $s$ is its uncomputability \cite{chaitin}. In other words, there is no program which takes a string $s$ as input and produces the integer $K(s)$ as output. This is usually considered a major problem, but one ought to expect a universal measure of complexity to have such a property. On the other hand, $K$ is more precisely upper semi-computable, meaning that one can find upper bounds, as we will do by applying a technique based on another semi-computable measure to be presented in the next section. 

The invariance theorem guarantees that complexity values will only diverge by a constant $c$ (e.g. the length of a compiler, a translation program between $U_1$ and $U_2$) and that they will converge at the limit. \\

\noindent \textbf{Invariance Theorem} (\cite{calude,li}): If $U_1$ and $U_2$ are two universal Turing machines and $K_{U_1}(s)$ and $K_{U_2}(s)$ the algorithmic complexity of $s$ for $U_1$ and $U_2$, there exists a constant $c$ such that:  

\begin{equation}
\label{invariance}
| K_{U_1}(s) - K_{U_2}(s) | < c
\end{equation} 

Hence the longer the string, the less important $c$ is (i.e. the choice of programming language or universal Turing machine). However, in practice $c$ can be arbitrarily large because the invariance theorem tells nothing about the rate of convergence between $K_{U_1}$ and $K_{U_2}$ for a string $s$ of increasing length, thus having an important impact on short strings. 

%Lossless compression algorithms have been used to approximate the Kolmogorov complexity of an object (e.g. a string). However, if the string is shorter than, for example, the size of the decompression algorithm, there will not be a way to compress the string into something shorter still. The result will be so dependent on the size of the decompression algorithm that the final value of the compressed length will be too unstable under different lossless compression/decompression algorithms.

\section{Solomonoff-Levin Algorithmic Probability}
\label{sec:coding}

The algorithmic probability (also known as Levin's semi-measure) of a string $s$ is a measure that describes the expected probability of a random program $p$ running on a universal (prefix-free\footnote{The group of valid programs forms a prefix-free set (no element is a prefix of any other, a property necessary to keep $0 < m(s) < 1$.) For details see \cite{calude}.}) Turing machine $T$ producing $s$ upon halting. Formally  \cite{solomonoff,levin,chaitin}, 

\begin{equation}
\label{coding}
m(s) = \sum_{p:T(p) = s} 1/2^{|p|}
\end{equation}
%That is, the sum over all the programs for which $T$ with $p$ outputs $s$ and halts. 

Levin's semi-measure\footnote{It is called a \emph{semi} measure because the sum is never 1, unlike probability measures. This is due to the Turing machines that never halt.} $m(s)$ defines a distribution known as the Universal Distribution (a beautiful introduction is given in \cite{kircher}). It is important to notice that the value of $m(s)$ is dominated by the length of the smallest program $p$ (when the denominator is larger). However, the length of the smallest $p$ that produces the string $s$ is $K(s)$. The semi-measure $m(s)$ is therefore also uncomputable, because for every $s$, $m(s)$ requires the calculation of $2^{-K(s)}$, involving $K$, which is itself uncomputable. An alternative to the traditional use of compression algorithms is the use of the concept of algorithmic probability to calculate $K(s)$ by means of the following theorem.\\

\noindent \textbf{Coding Theorem} (Levin~\cite{levin}):
\begin{equation}
|-\log_2 m(s) - K(s)| < c
\end{equation}

This means that if a string has many descriptions it also has a short one. It beautifully connects frequency to complexity, more specifically the frequency of occurrence of a string with its algorithmic (Kolmogorov) complexity. The Coding theorem implies that \cite{cover,calude} one can calculate the Kolmogorov complexity of a string from its frequency \cite{zenil2007,delahaye2007,thesis,delahayezenil}, simply rewriting the formula as:

\begin{equation}
K_m(s)=-\log_2 m(s) + O(1)
\end{equation}

%That means that $K$ and $m$ are in perfect anti-correlation and after all even if they may serve for different purposes they are essentially the same complexity measure (unlike, for example, Bennett's Logical Depth \cite{bennett} as we have shown in \cite{numerical}). 
An important property of $m$ as a semi-measure is that it dominates any other effective semi-measure $\mu$, because there is a constant $c_\mu$ such that for all $s$, $m(s) \geq c_\mu\mu(s)$. For this reason $m(s)$ is often called a \emph{Universal Distribution}  \cite{kircher}.

\section{The Coding Theorem Method}
\label{D}

%One can attempt to approximate $m(s)$ by running every Turing machine following a particular enumeration. A ``natural" one is a quasi-lexicographical order from shorter to longer machine length in number of $n$ states and $m$ symbols (denoted by $(n, m)$) and therefore in number of transition rules. It is clear that in this fashion once a machine produces $s$ for the first time, one can directly calculate an exact value of $K$. Because this is the length of the first Turing machine in the enumeration of programs of increasing size that produces $s$, there is no shorter machine producing $s$, and based on Turing universality we know there is a machine $T\in(n, m)$ that produces $s$. 

Let $D(n, m)$ be a function~\cite{delahayezenil} defined as follows:

%{ T \in (n,m) | U(T)=s}/ {T\in (n,m) |  U(T) halts}

\begin{equation}
\label{Deq}
D(n, m)(s)=\frac{|\{T\in(n, m)\ :\ T \textit{ produces } s\}|}{|\{T\in(n, m)
  \ :\ T \textit{ halts }\}|} 
\end{equation}

\noindent Where $(n,m)$ denotes the set of Turing machines with
$n$ states and $m$ symbols, running with empty input, and
$|A|$ is, in this case, the cardinality of the set
$A$. In~\cite{thesis,delahayezenil} we calculated the output
distribution of Turing machines with 2-symbols and $n=1, \ldots, 4$
states for which the Busy Beaver \cite{rado} values are known, in
order to determine the halting time, and in~\cite{d5} results were
improved in terms of number and Turing machine size (5 states) and in
the way in which an alternative to the Busy Beaver information was
proposed, hence no longer needing exact information of halting times
in order to approximate an informative distribution. 
%That is, a total of 36, 10\,000, 7\,529\,536 and 11\,019\,960\,576 Turing machines respectively. 

%We have claimed before that because there are a large enough number of machines to run even for a small number of machine states ($n$), obtaining a Universal Distribution and applying the Coding theorem provides an alternative to the evaluation of $K(s)$ based on the frequency of production of Turing machines. 

Here we consider an experiment with 2-dimensional deterministic Turing machines (also called \textit{Turmites}) in order to estimate the Kolmogorov complexity of 2-dimensional objects, such as images that can represent space-time diagrams of simple systems. A \textit{Turmite} is a Turing machine which has an orientation and operates on a grid for ``tape". The machine can move in 4 directions rather than in the traditional left and right movements of a traditional Turing machine head. A reference to this kind of investigation and definition of 2D Turing machines can be found in~\cite{wolfram}, one popular and possibly one of the first examples of this variation of a Turing machine is Lagton's ant~\cite{langton} also proven to be capable of Turing-universal computation.

In Section~\ref{sec:strings-lenghts-10}, we will use the so-called \emph{Turmites} to provide evidence that Kolmogorov complexity evaluated through algorithmic probability is consistent with the other (and today only) method for approximating $K$, namely lossless compression algorithms. We will do this in an artful way, given that compression algorithms are unable to compress strings that are too short, which are the strings covered by our method. This will involve concatenating strings for which our method establishes a Kolmogorov complexity, which then are given to a lossless compression algorithm in order to determine whether it provides consistent estimations, that is, to determine whether strings are less compressible where our method says that they have greater Kolmogorov complexity and whether strings are more compressible where our method says they have lower Kolmogorov complexity. We provide evidence that this is actually the case. 

In Section~\ref{eca} we will apply the results from the Coding theorem method to approximate the Kolmogorov complexity of 2-dimensional evolutions of 1-dimensional, closest neighbor Cellular Automata as defined in \cite{wolfram}, and by way of offering a contrast to the approximation provided by a general lossless compression algorithm (Deflate). As we will see, in all these experiments we provide evidence that the method is just as successful as compression algorithms, but unlike the latter, it can deal with short strings.

\subsection{Deterministic 2-dimensional Turing machines (Turmites)}
\label{sec:two-dimens-turing}

Turmites or 2-dimensional (2D) Turing machines run not on a 1-dimensional tape but in a 2-dimensional unbounded grid or array. At each step they can move in four different directions (\emph{up}, \emph{down}, \emph{left}, \emph{right}) or \emph{stop}. Transitions have the format $\{n_1,m_1\} \to \{n_2,m_2,d\}$, meaning that when the machine is in state $n_1$ and reads symbols $m_1$, it writes $m_2$, changes to state $n_2$ and moves to a contiguous cell following direction $d$. If $n_2$ is the halting state then $d$ is $stop$. In other cases, $d$ can be any of the other four directions.

Let $(n,m)_{2D}$ be the set of Turing machines with $n$ states and $m$ symbols. These machines have $n m$ entries in the transition table, and for each entry $\{n_1,m_1\}$ there are $4nm+m$ possible instructions, that is, $m$ different halting instructions (writing one of the different symbols) and $4nm$ non-halting instructions (4 directions, $n$ states and $m$ different symbols). So the number of machines in $(n,m)_{2D}$ is $(4nm+m)^{nm}$. It is possible to enumerate all these machines in the same way as 1D Turing machines (e.g. as has been done in \cite{wolfram} and \cite{joost}). We can assign one number to each entry in the transition table. These numbers go from 0 to $4nm+m-1$ (given that there are $4nm + m$ different instructions). The numbers corresponding to all entries in the transition table (irrespective of the convention followed in sorting them) form a number with $nm$ digits in base $4nm + m$. Then, the translation of a transition table to a natural number and vice versa can be done through elementary arithmetical operations.

We take as output for a 2D Turing machine the minimal array that includes all cells visited by the machine. Note that this probably includes cells that have not been visited, but it is the more natural way of producing output with some regular format and at the same time reducing the set of different outputs. 

\begin{figure}[htbp!]
  \centering
  \begin{minipage}[c]{0.4\linewidth}
  \begin{tabular}[b]{l@{\mbox{}\ $\to$\ \mbox{}}l}
$\{1, 1\}$ & $\{0, 0, stop\}$ \\
$\{1, 0\}$ & $\{3, 1, right\}$ \\
$\{2, 1\}$ & $\{3, 1, up\}$ \\
$\{2, 0\}$ & $\{0, 1, stop\}$ \\
$\{3, 1\}$ & $\{0, 0, down\}$ \\
$\{3, 0\}$ & $\{2, 0, left\}$
\end{tabular}  
  \end{minipage}
  \ \ 
  \begin{minipage}[c]{0.5\linewidth}
    \includegraphics[width=6.5cm]{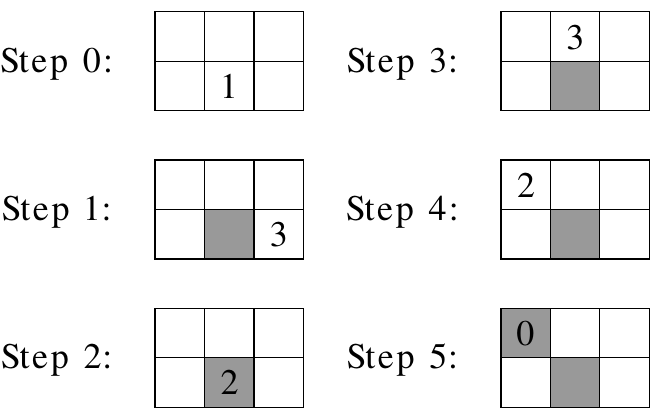}
  \vspace{1cm}  
  \end{minipage}
  \noindent\rule{6cm}{0.4pt}\\
  
  \vspace{1cm}
  \includegraphics[width=9.5cm]{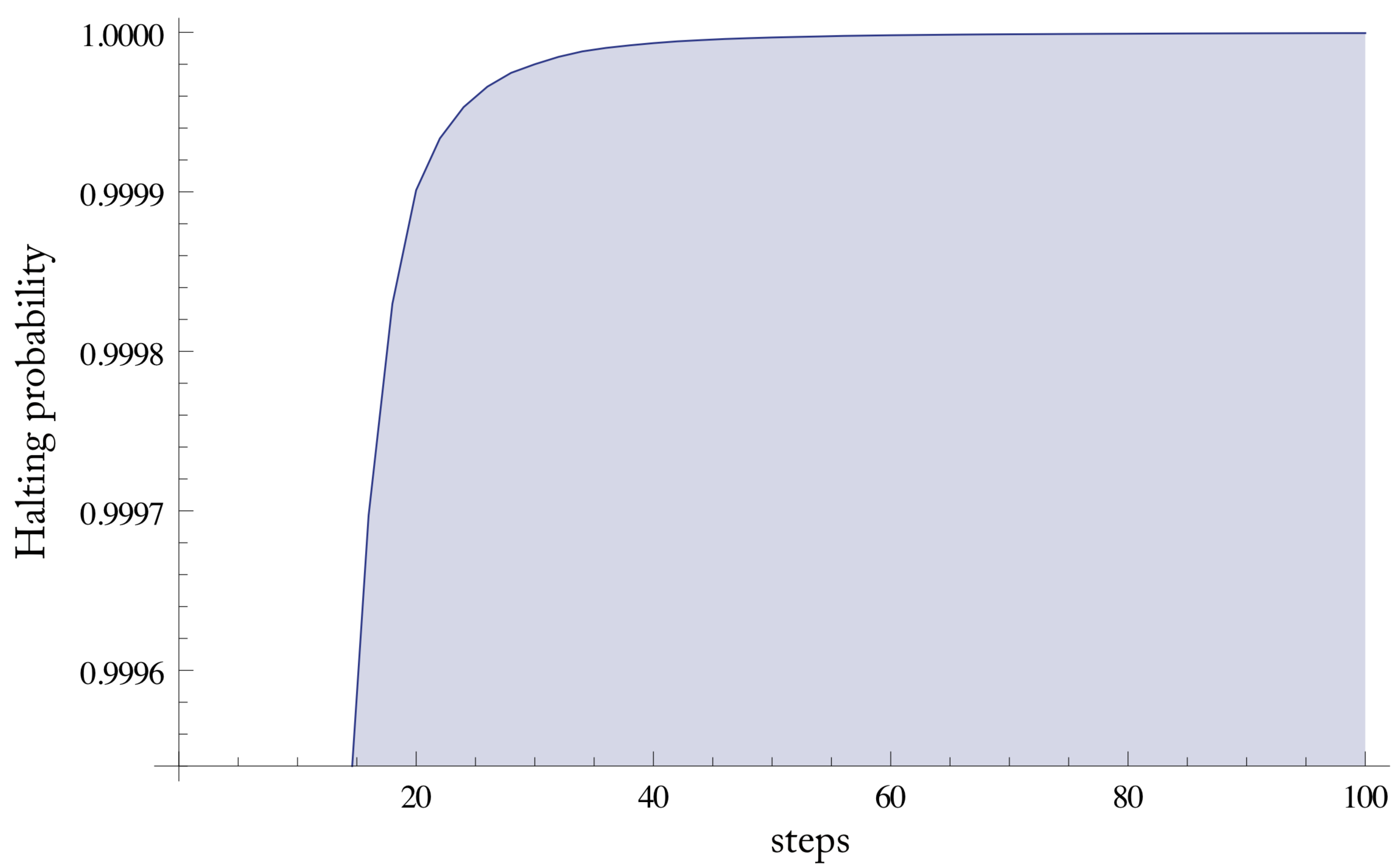}
  \caption{Top: Example of a deterministic 2-dimensional Turing machine. Bottom: Accumulated runtime distribution for $(4,2)_{2D}$.}
  \label{fig:exam2D32}
\end{figure}

Fig.~\ref{fig:exam2D32} shows an example of the transition table of a Turing machine in $(3,2)_{2D}$ and its execution over a `0'-filled grid. We show the portion of the grid that is returned as the output array. Two of the six cells have not been visited by the machine.

\section{An approximation to the Universal Distribution}
\label{sec:experiment}

We have run all machines in $(4,2)_{2D}$ just as we have done before
for deterministic 1-dimensional Turing machines~\cite{delahayezenil,d5}. That is, considering the output of all different machines starting both in a `0'-filled grid (all white) and in a
`1'-filled (all black) grid. Symmetries are described and used in the same way than in~\cite{d5} in order to avoid running a larger number of machines whose output can be predicted from other equivalent machines (by rotation, transposition, 1-complementation, reversion, etc.) that produce equivalent outputs with the same frequency.

We also used a reduced enumeration to avoid running certain trivial machines whose behavior can be predicted from the transition table, as well as filters to detect non-halting machines before exhausting the entire runtime. In the reduced enumeration we considered only machines with an initial transition moving to the right and changing to a different state than the initial and halting states. Machines moving to the initial state at the starting transition run forever, and machines moving to the halting state produce single-character output. So we reduce the number of initial transitions in $(n,m)_{2D}$ to $m(n-1)$ (the machine can write any of the $m$ symbols and change to any state in $\{2,\cdots,n\}$). The set of different machines is reduced accordingly to $k(n-1)(4nm+m)^{nm-1}$. To enumerate these machines we construct a mixed-radix number, given that the digit corresponding to the initial transition now goes from 0 to $m(n-1)-1$. To the output obtained when running this reduced enumeration we add the single-character arrays that correspond to machines moving to the initial state at the starting transition. These machines and their output can be easily quantified. Also, to take into account machines with the initial transition moving in a different direction than the right one, we consider the 90, 180 and 270 degree rotations of the strings produced, given that for any machine moving up (left/down) at the initial transition, there is another one moving right that produces the identical output but rotates -90 (-180/-270) degrees.

\subsection{Setting the runtime}
\label{sec:setting-runtime}

The Busy Beaver runtime value for $(4,2)$ is 107 steps before halting. But no equivalent Busy Beavers are known for 2-dimensional Turing machines (although variations of Turmite's Busy Beaver functions have been proposed~\cite{pegg}). So to set the runtime in our experiment we generated a sample of $334\times 10^8$ random machines in the reduced enumeration. We used a runtime of 2000 steps for the runtime sample, this is 10.6\% of the machines in the reduced enumeration for $(4,2)_{2D}$, but 1500 steps for running all $(4,2)_2D$. These machines were generated instruction by instruction. As we have explained above, it is possible to assign a natural number to every instruction. So to generate a random machine in the reduced enumeration for $(n,m)_{2D}$ we produce a random number from 0 to $m(n-1)-1$ for the initial transition and from 0 to $4nm+m-1$ for the other $nm-1$ transitions. We used the implementation of the Mersenne Twister in
the Boost C++ library. The output of this sample was the distribution of the runtime of the halting machines. 

Fig.~\ref{fig:exam2D32} shows the probability that a random halting machine will halt in at most the number of steps indicated on the horizontal axis. For 100 steps this probability is $0.9999995273$. Note that the machines in the sample are in the reduced enumeration, a large number of very trivial machines halting in just one step having been removed. So in the complete enumeration the probability of halting in at most 100 steps is even greater. 

But we found some high runtime values---precisely 23 machines required more than 1000 steps. The highest value was a machine progressing through 1483 steps before halting. So we have enough evidence to believe that by setting the runtime at 2000 steps we have obtained almost all (if not all) output arrays. We ran all $6 \times 34^7$ Turing machines in the reduced enumeration for $(4,2)_{2D}$. Then we applied the completions explained before.

\section{Output Analysis}
\label{sec:results}

The final output represents the result of $2(4nm+m)^2$ executions (all machines in $(4,2)_{2D}$ starting with both blank symbols `0' and `1'). We found $3\,079\,179\,980\,224$ non-halting machines and $492\,407\,829\,568$ halting machines. A number of $1\,068\,618$ different binary arrays were produced after 12 days of calculation with a supercomputer of medium size (a 25 x86-64 CPUs running at 2128 MHz each with 4 GB of memory each, located at the Centro Inform\'atico Cient\'ifico de Andaluc\'ia (CICA), Spain.

\begin{figure}[htbp!]
  \centering
  \includegraphics[width=12.5cm]{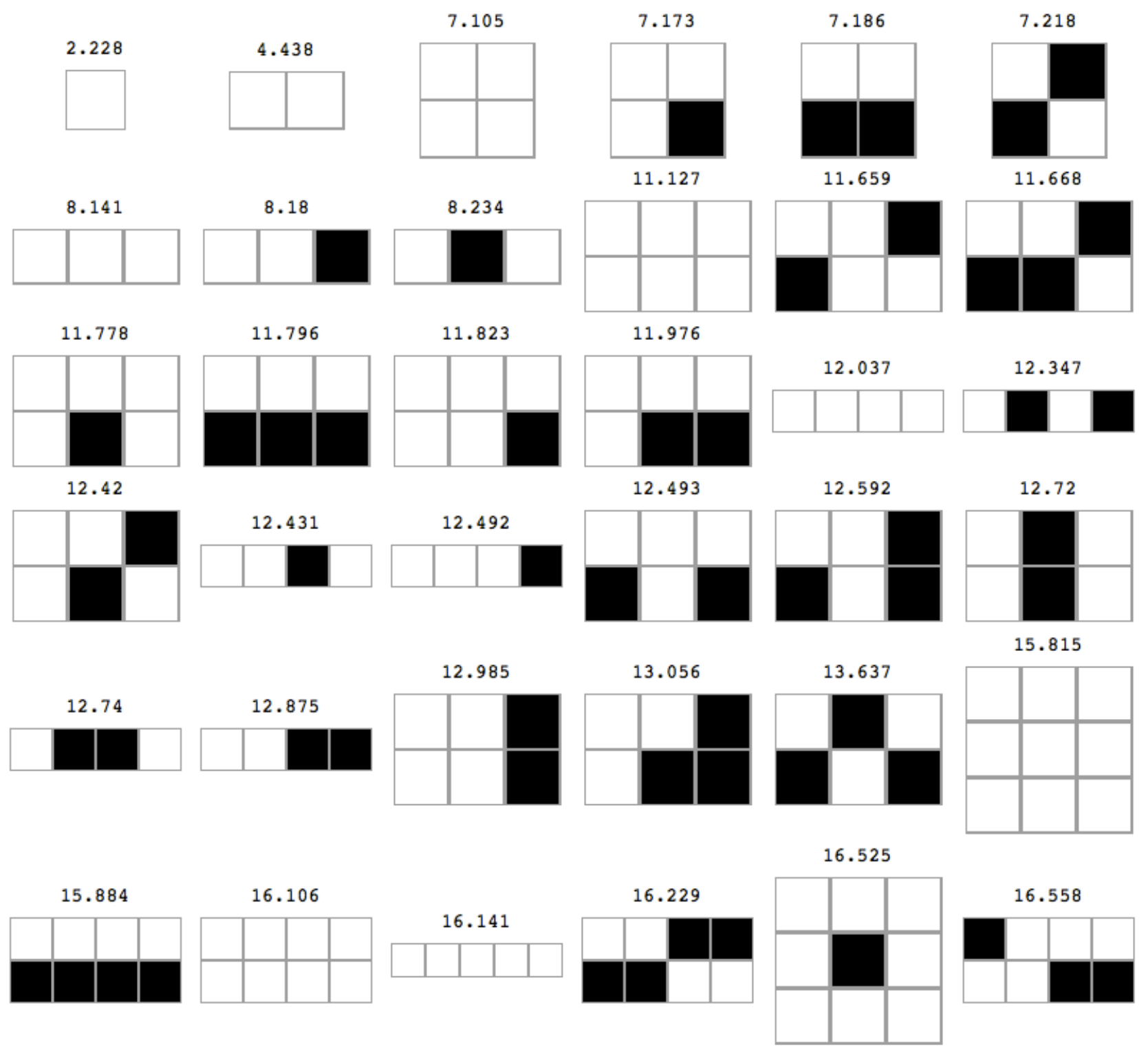}
  \caption{The top 36 objects in $D(4,2)_{2D}$ preceded by their $K_{m,2D}$ values, sorted by higher to lower frequency and therefore from smaller to larger Kolmogorov complexity after application of the Coding theorem). Only non-symmetrical cases are displayed. The grid is only for illustration purposes.}
  \label{top}
\end{figure}

Let $D(4,2)_{2D}$ be the set constructed by dividing the occurrences of each different array by the number of halting machines as a natural extension of Eq.~\ref{Deq} for 2-dimensional Turing machines. Then, for every string $s$, 

\begin{equation}
K_{m,2D}(s) = -\log_2(D(4,2)(s))
\end{equation}
\noindent using the Coding theorem (Eq.~\ref{coding}). Fig.~\ref{top} shows the top 36 objects in $D(4,2)_{2D}$, that is the objects with lowest Kolmogorov complexity values.

\subsection{Evaluating 2-dimensional Kolmogorov complexity}
\label{sec:glance-into-k_m}

\begin{figure}[htbp]
\begin{center}
 \includegraphics[width=12cm]{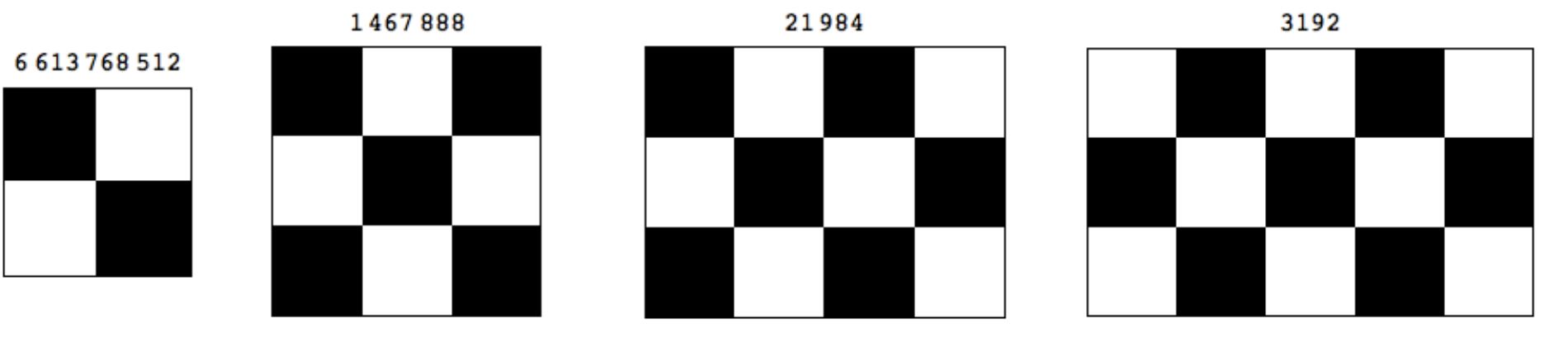}\\

 \vspace{1cm}
 
 \noindent\rule{6cm}{0.4pt}\\
 
 \vspace{1cm}
  \includegraphics[width=12.4cm]{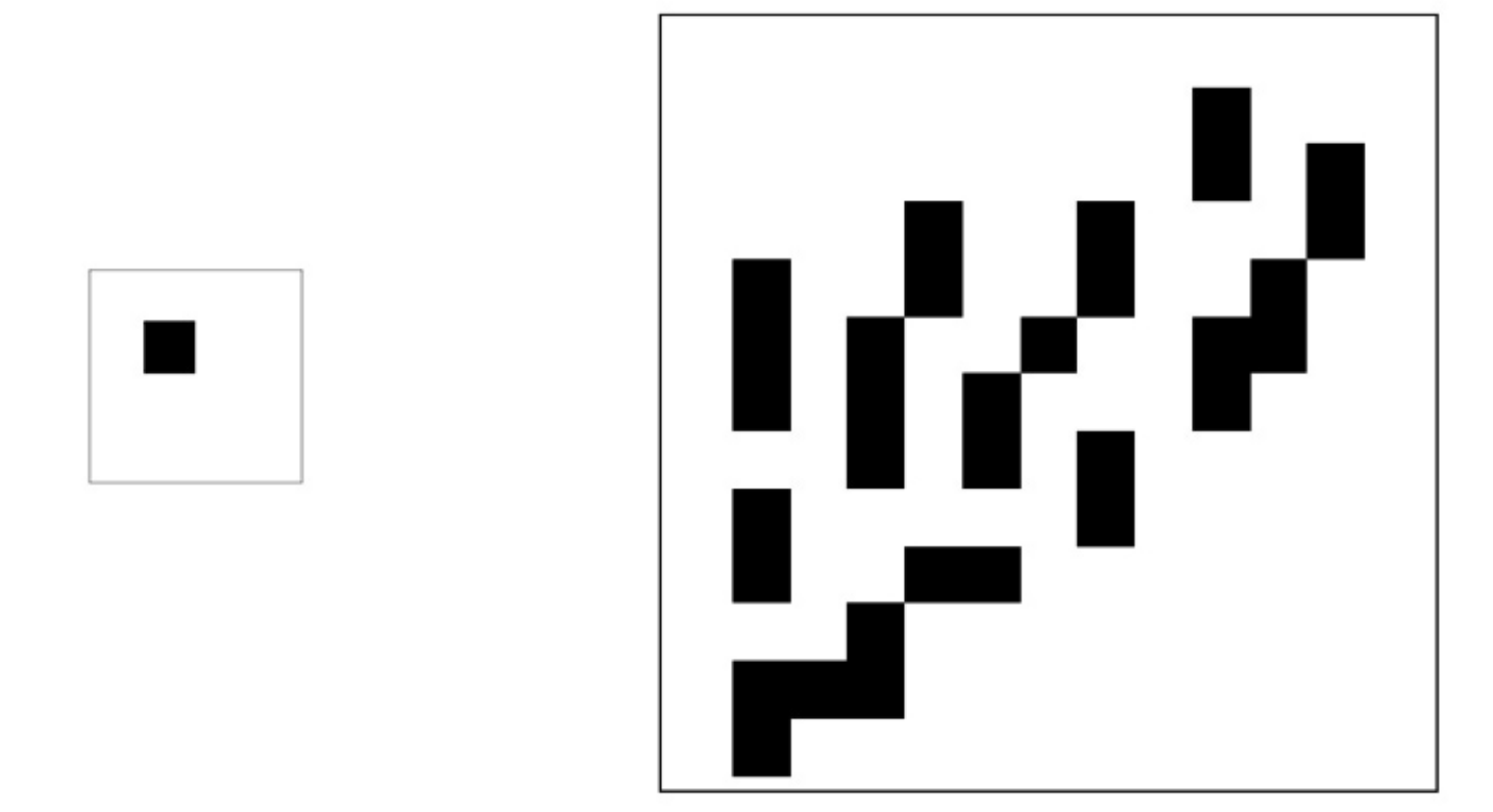}
\caption{Top: Frequency of appearance of symmetric ``checkerboard" patterns sorted from more to less frequent according to $D(4,2)_{2D}$ (displayed only non-symmetrical cases under rotation and complementation). The checkerboard of size $4\times4$ doesn't occur. However, all $3\times3$ as seen in Fig.~\ref{fig:arrays3x3}, including the ``checkerboard" pattern of size $3\times3$ do occur. Bottom: Symmetry breaking from a fully deterministic set of symmetric computational rules. Bottom Left: With a value of $K_{m,2D}=6.7$ this is the simplest $4\times4$ square array after the preceding all-blank $4\times4$ array (with $K_{m,2D}=6.4$) and before the $4\times4$ square array with a black cell in one of the array corners (with complexity $K_{m,2D}=6.9$). Bottom Right:  The only and most complex square array (with 15 other symmetrical cases) in $D(4,2)_{2D}$ with $K_{m,2D}=34.2561$. Another way to see this array is as one among those of length 13 with low complexity given that it occurred once in the sampled distribution in the classification unlike all other square arrays of the same size that are missing in $D(4,2)_{2D}$.}
\label{checkerboard}
\end{center}
\end{figure}

$D(4,2)_{2D}$ denotes the frequency distribution (a calculated Universal Distribution) from the output of deterministic 2-dimensional Turing machines, with associated complexity measure $K_{m,2D}$. $D(4,2)_{2D}$ distributes 1\,068\,618 arrays into 1272 different complexity values, with a minimum complexity value of 2.22882 bits (an explanation of non-integer program-size complexity is given in \cite{d5} and \cite{numerical}), a maximum value of 36.2561 bits and a mean of 35.1201. Considering the number of possible square binary arrays given by the formula $2^{d\times d}$ (without considering any symmetries), $D(4,2)_{2D}$ can be said to produce all square binary arrays of length up to $3 \times 3$, that is $\sum_{d=1}^3 2^{d\times d}=530$ square arrays, and 60016 of the $2^{(4\times 4)}=65536$ square arrays with side of length (or dimension) $d=4$. It only produces 84104 of the 33\,554\,432 possible square binary arrays of length $d=5$ and only 11328 of the possible 68\,719\,476\,736 of dimension $d=6$. The largest square array produced in $D(4,2)_{2D}$ is of side length $d=13$ (Left of Fig.~\ref{checkerboard}) out of a possible $748\times10^{48}$; it has a $K_{m,2D}$ value equal to 34.2561.

%\begin{figure}[htbp]
%\begin{center}
% \includegraphics[width=8.5cm]{lowclimbers.pdf}
%\caption{With a value of $K_{m,2D}=6.7$ this is the greatest $4\times4$ square array after the preceding all-blank $4\times4$ array (with $K_{m,2D}=6.4$) and before the $4\times4$ square array with a black cell in one of the corners (with complexity $K_{m,2D}=6.9$).}
%\label{squareclimbers}
%\end{center}
%\end{figure}

\begin{figure}[htbp]
\begin{center}
 \includegraphics[width=12cm]{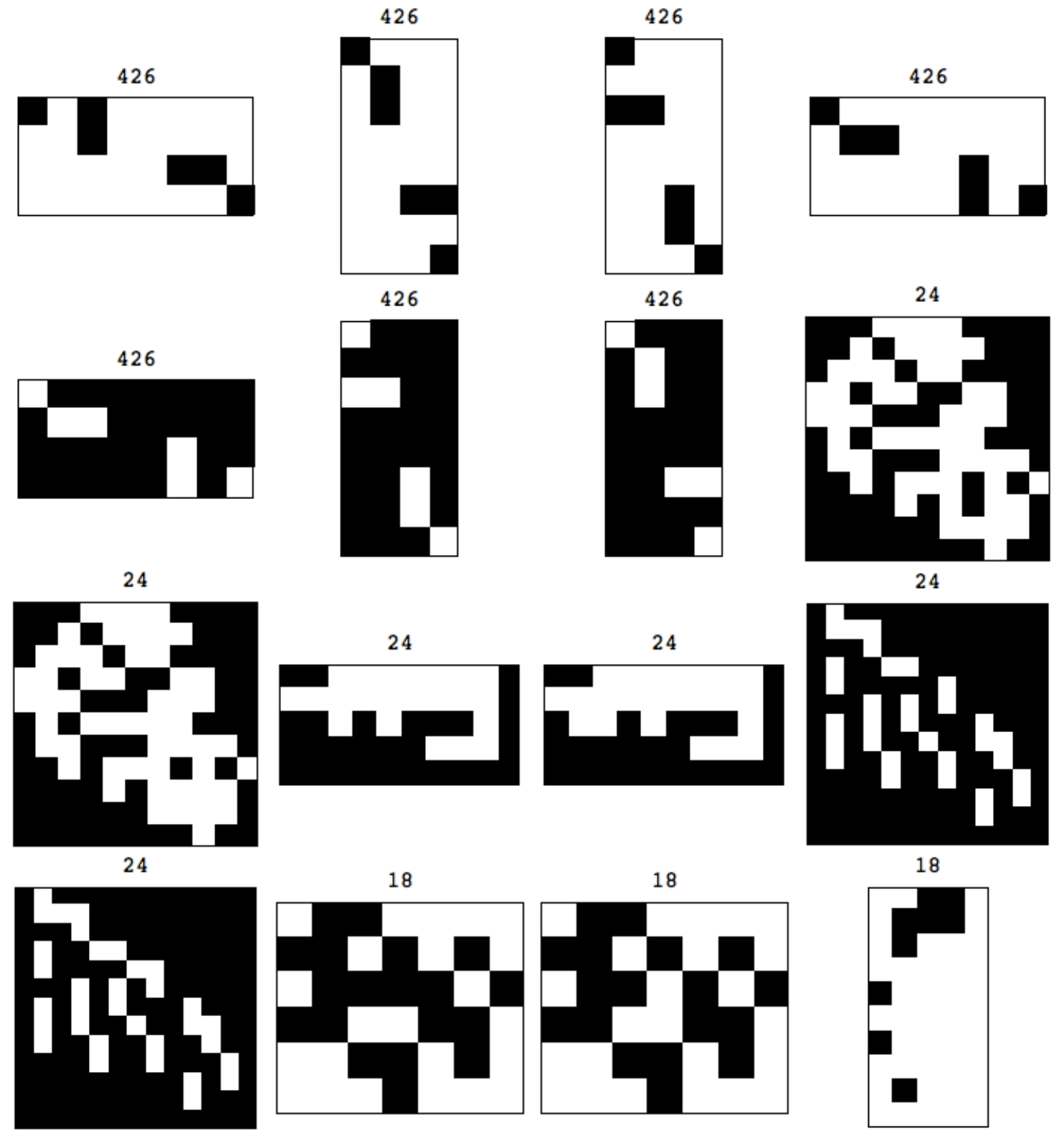}
\caption{Symmetry breaking from fully deterministic symmetric computational rules. Bottom 16 objects in the classification with lowest frequency, or being most random according to $D(4,2)_{2D}$. It is interesting to note the strong similarities given that similar-looking cases are not always exact symmetries. The arrays are preceded by the number of occurrences of production from all the $(4,2)_{2D}$ Turing machines.}
\label{bottom}
\end{center}
\end{figure}

What one would expect from a distribution where simple patterns are more frequent (and therefore have lower Kolmogorov complexity after application of the Coding theorem) would be to see patterns of the ``checkerboard" type with high frequency and low random complexity ($K$), and this is exactly what we found (see Fig.~\ref{checkerboard}), while random looking patterns were found at the bottom among the least frequent ones (Fig.~\ref{bottom}). 

%\begin{figure}[htbp]
%\begin{center}
 %\includegraphics[width=10cm]{squareclimbers.pdf}
%\caption{Other ``climber" arrays occurring in $D(4,2)_{2D}$. That is, arrays that appear significantly before (less random) in the classification than the rest of the arrays of about the same size.}
%\label{squareclimbers2}
%\end{center}
%\end{figure}

We have coined the informal notion of a ``climber" as an object in the frequency classification (from greatest to lowest frequency) that appears better classified among objects of smaller size rather than with the arrays of their size, this is in order to highlight possible candidates for low complexity, hence illustrating how the process make low complexity patterns to emerge. For example, ``checkerboard" patterns (see Fig.~\ref{checkerboard}) seem to be natural ``climbers" because they come significantly early (more frequent) in the classification than most of the square arrays of the same size. In fact, the larger the checkerboard array, the more of a climber it seems to be. This is in agreement with what we have found in the case of strings \cite{thesis,delahayezenil,d5} where patterned objects \emph{emerge} (e.g. (01)$^n$, that is, the string 01 repeated $n$ times), appearing relatively increasingly higher in the frequency classifications the larger $n$ is, in agreement with the expectation that patterned objects should also have low Kolmogorov complexity.

\begin{figure}[htbp]
\begin{center}
 \includegraphics[width=13cm]{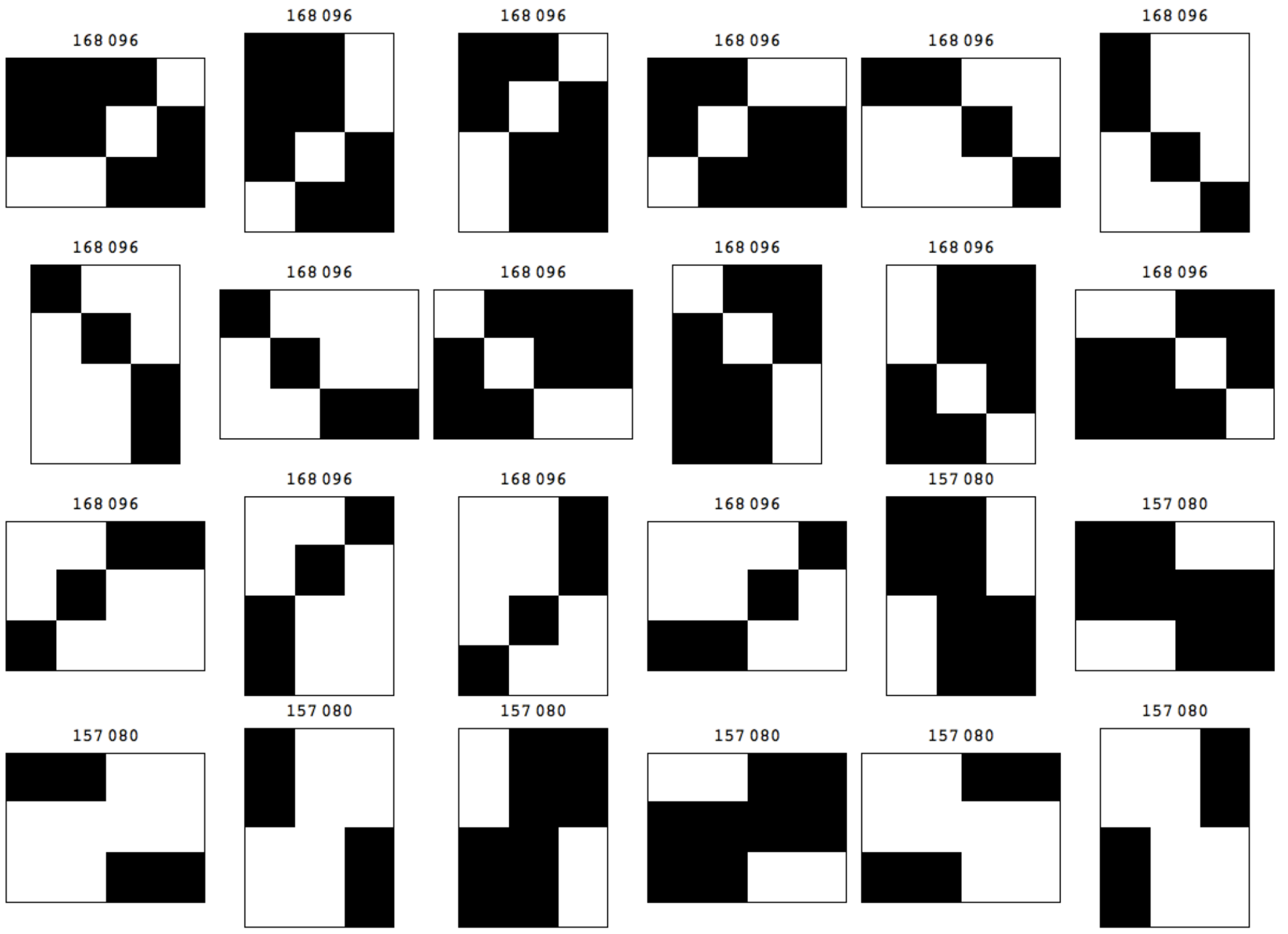}
\caption{Two ``climbers" (and all their symmetric cases) found in $D(4,2)_{2D}$. Symmetric objects have higher frequency and therefore lower Kolmogorov complexity. Nevertheless, a fully deterministic algorithmic process starting from completely symmetric rules produces a range of patterns of high complexity and low symmetry.}
\label{climbers}
\end{center}
\end{figure}

An attempt of a definition of a climber is a pattern $P$ of size $a\times b$ with small complexity among all $a\times b$ patterns, such that there exists smaller patterns $Q$ (say $c\times d$, with $cd<ab$) such that $K_m(P)<K_m(Q)<median(K_m(\textit{all $ab$ patterns}))$.

%\begin{figure}[htbp]
%\begin{center}
 %\includegraphics[width=8.5cm]{squaredist.pdf}
%\caption{Distribution of square binary arrays. $(4,2)_{2D}$ was divided into 5 parts with F1 the first fifth, F2 the second fifth, and so on.}
%\label{squaredist}
%\end{center}
%\end{figure}

For example, Fig.~\ref{climbers} shows arrays that come together among groups of much shorter arrays, thereby demonstrating, as expected from a measure of randomness, that array---or string---size is not what determines complexity (as we have shown before in \cite{thesis,delahayezenil,d5} for binary strings). The fact that square arrays may have low Kolmogorov complexity can be understood in several ways, some of which strengthen the intuition that square arrays should be less Kolmogorov random, such as for example, the fact that for square arrays one only needs the information of one of its dimensions to determine the other, either height or width.

%In Fig.~\ref{squareclimbers2} more low complexity arrays are depicted, coming together in a block in the frequency distribution before the larger group of arrays of the same size (hence being what we call ``climbers"). 

Fig.~\ref{climbers} shows cases in which square arrays are significantly better classified towards the top than arrays of similar size. Indeed, 100\% of the squares of size $2 \times 2$ are in the first fifth (F1), as are the $3 \times 3$ arrays. Square arrays of $4\times4$ are distributed as follows when dividing $(4,2)_{2D}$ in 5 equal parts: 72.66\%, 15.07\%, 6.17359\%, 2.52\%, 3.56\%.

%Fig.~\ref{squaredist} shows the distribution of square arrays. The 3 rings in Fig.~\ref{squaredist} show how square arrays of length $1 < d < 5$ in $(4,2)_{2D}$, for which $(4,2)_{2D}$ contains most $2^{d\times d}$ cases, are distributed.

\begin{figure}[htbp!]
  \centering
  \includegraphics[width=10.5cm]
{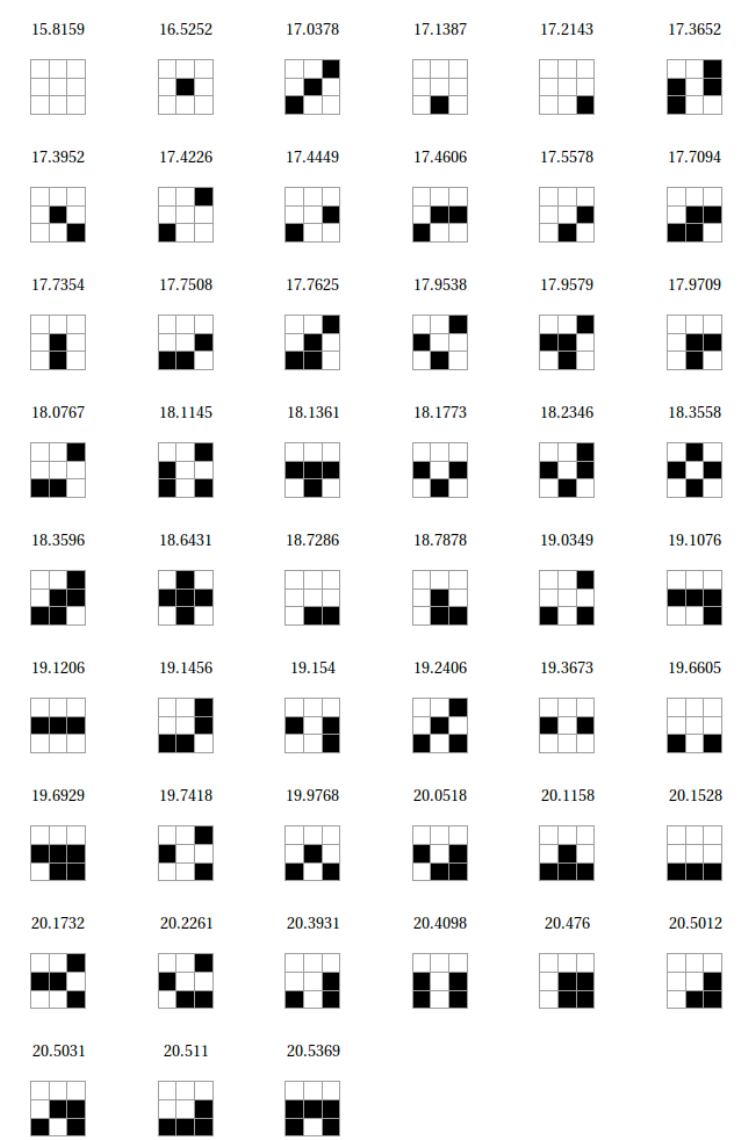}
  \caption{Complete reduced set (non-symmetrical cases under reversion and complementation) of $3\times 3$ patches in $K_{m,2D}$ sorted from lowest to greatest Kolmogorov complexity after application of the Coding theorem (Eq.~\ref{coding}) to the output frequency of 2-D Turing machines. We denote this set by $K_{m,2D_{3\times3}}$. For example, the 2 glider configurations in the Game of Life~\cite{life} come with high Kolmogorov complexity (with approximated values of 20.2261 and 20.5031).}
  \label{fig:arrays3x3}
\end{figure}

\section{Validation of the Coding Theorem Method by Compressibility}
\label{comparison} 

One way to validate our method based on the Coding theorem (Eq.~\ref{coding}) is to attempt to measure its departure from the compressibility approach. This cannot be done directly, for as we have explained, compression algorithms perform poorly on short strings, but we did find a way to partially circumvent this problem by selecting subsets of strings for which our \emph{Coding theorem method} calculated a high or low complexity which were then used to generate a file of length long enough to be compressed.

\subsection{Comparison of $K_m$ and approaches based on compression}
\label{sec:strings-lenghts-10}

It is also not uncommon to detect instabilities in the values retrieved by a compression algorithm for short strings, as explained in Section~\ref{compress}, strings which the compression algorithm may or may not compress. This is not a malfunction of a particular lossless compression algorithm (e.g. Deflate, used in most popular computer formats such as ZIP and PNG) or its implementation, but a commonly encountered problem when lossless compression algorithms attempt to compress short strings. 

When researchers have chosen to use compression algorithms for reasonably long strings, they have proven to be of great value, for example, for DNA false positive repeat sequence detection in genetic sequence analysis \cite{rivals}, in distance measures and classification methods \cite{cilibrasi}, and in numerous other applications \cite{li}. However, this effort has been hamstrung by the limitations of compression algorithms--currently the only method used to approximate the Kolmogorov complexity of a string--given that this measure is not computable. 

In this section we study the relation between $K_m$ and approaches to Kolmogorov complexity based on compression. We show that both approaches are consistent, that is, strings with higher $K_m$ value are less compressible than strings with lower values. This is as much validation of $K_m$ and our Coding theorem method as it is for the traditional lossless compression method as approximation techniques to Kolmogorov complexity. The Coding theorem method is, however, especially useful for short strings where losses compression algorithms fail, and the compression method is especially useful where the Coding theorem is too expensive to apply (long strings).

\subsubsection{Compressing strings of length 10 to 15}

\begin{figure}[htbp!]
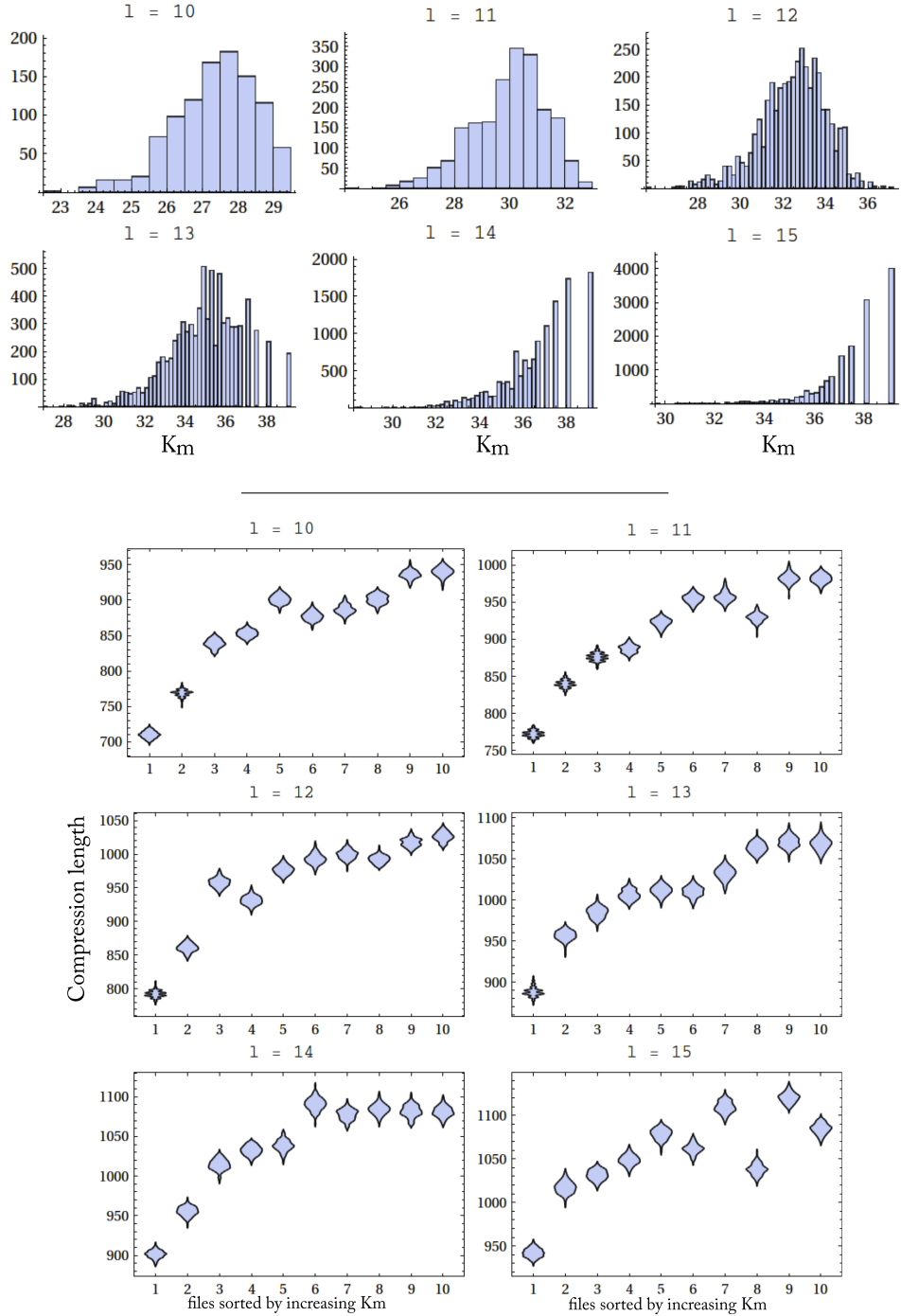

  \centering
  \vspace{-.5cm}
  \includegraphics[width=12.5cm]{Fig9.pdf}
\noindent\rule{6cm}{0.4pt}\\

\vspace{.3cm}

      \includegraphics[width=11cm]{Fig10.pdf}
  \caption{Top: Distribution of complexity values for different string lengths ($l$). Bottom: Distribution of the compressed lengths of the files.}\label{fig:dislen10to15}
\end{figure}

For this experiment we have selected the strings in $D(5)$ with lengths ranging from 10 to 15. $D(5)$ is the frequency distribution of strings produced by all 1-dimensional deterministic Turing machines as described in \cite{d5}. Table~\ref{tab:numberS10to15} shows the number of $D(5)$ strings with these lengths. Up to length 13 we have almost all possible strings. For length 14 we have a considerable number and for length 15 there are less than $50\%$ of the $2^{15}$ possible strings. The distribution of complexities is shown in Fig.~\ref{fig:dislen10to15}. 

\begin{table}[htbp!]
  \centering
\begin{tabular}{cc}\hline
Length ($l$) & Strings\\\hline
  10 & 1024 \\
  11 & 2048 \\
  12 & 4094 \\
  13 & 8056 \\
  14 & 13068 \\
  15 & 14634 \\\hline
\end{tabular}    
  \caption{Number of strings of length 10 to 15 found in $D(5)$}
  \label{tab:numberS10to15}
\end{table}

As expected, the longer the strings, the greater their average complexity. The overlapping of strings with different lengths that have the same complexity correspond to climbers. The experiment consisted in creating files with strings of different $K_m$-complexity but equal length (Files with more complex (random) strings are expected to be less compressible than files with less complex (random) strings). This was done in the following way. For each $l$ ($10\leq l\leq 15$), we let $S(l)$ denote the list of strings of length $l$, sorted by increasing $K_m$ complexity. For each $S(l)$ we made a partition of 10 sets with the same number of consecutive strings. Let's call these partitions $P(l,p)$, $1\leq p\leq 10$.

Then for each $P(l,p)$ we have created 100 files, each with 100 random strings in $P(l,p)$ in random order. We called these files $F(l,p,f)$, $1\leq f\leq 100$. Summarizing, we now have:

\begin{itemize}
\item 6 different string lengths $l$, from 10 to 15, and for each length
\item 10 partitions (sorted by increasing complexity) of the strings with length $l$, and 
\item 100 files with 100 random strings in each partition.
\end{itemize}

This makes for a total of 6\,000 different files. Each file contains 100 different binary strings, hence with length of $100\times l$ symbols. 

A crucial step is to replace the binary encoding of the files by a larger alphabet, retaining the internal structure of each string. If we compressed the files $F(l,p,f)$ by using binary encoding then the final size of the resulting compressed files would depend not only on the complexity of the separate strings but on the patterns that the compressor discovers along the whole file. To circumvent this we chose two different symbols to represent the `0' and `1' in each one of the 100 different strings in each file. The same set of 200 symbols was used for all files. We were interested in using the most standard symbols we possibly could, so we created all pairs of characters from `a' to `p' (256 different pairs) and from this set we selected 200 two-character symbols that were the same for all files. This way, though we do not completely avoid the possibility of the compressor finding patterns in whole files due 
to the repetition of the same single character in different strings, we considerably reduce the impact of this phenomenon. 

The files were compressed using the \emph{Mathematica} function \texttt{Compress}, which is an implementation of the Deflate algorithm (Lempel-Ziv plus Huffman coding). Fig.~\ref{fig:dislen10to15} shows the distributions of lengths of the compressed files for the different string lengths. The horizontal axis shows the 10 groups of files in
increasing $K_m$. As the complexity of the strings grows (right part of the diagrams), the compressed files are larger, so they are harder to compress. The relevant exception is length 15, but this is probably related to the low number of strings of that length that we have found, which  are surely not the most complex strings of length 15. 

We have used other compressors such as GZIP (which uses Lempel-Ziv algorithm LZ77) and BZIP2 (Burrows-Wheeler block sorting text compression algorithm and Huffman coding), with several compression levels. The results are similar to those shown in Fig.~\ref{fig:dislen10to15}.

\subsubsection{Comparing $(4,2)_{2D}$ and $(4,2)$}
\label{sec:comparing-2d-4}

\begin{figure}[htbp]
\begin{center}
 \includegraphics[width=12cm]{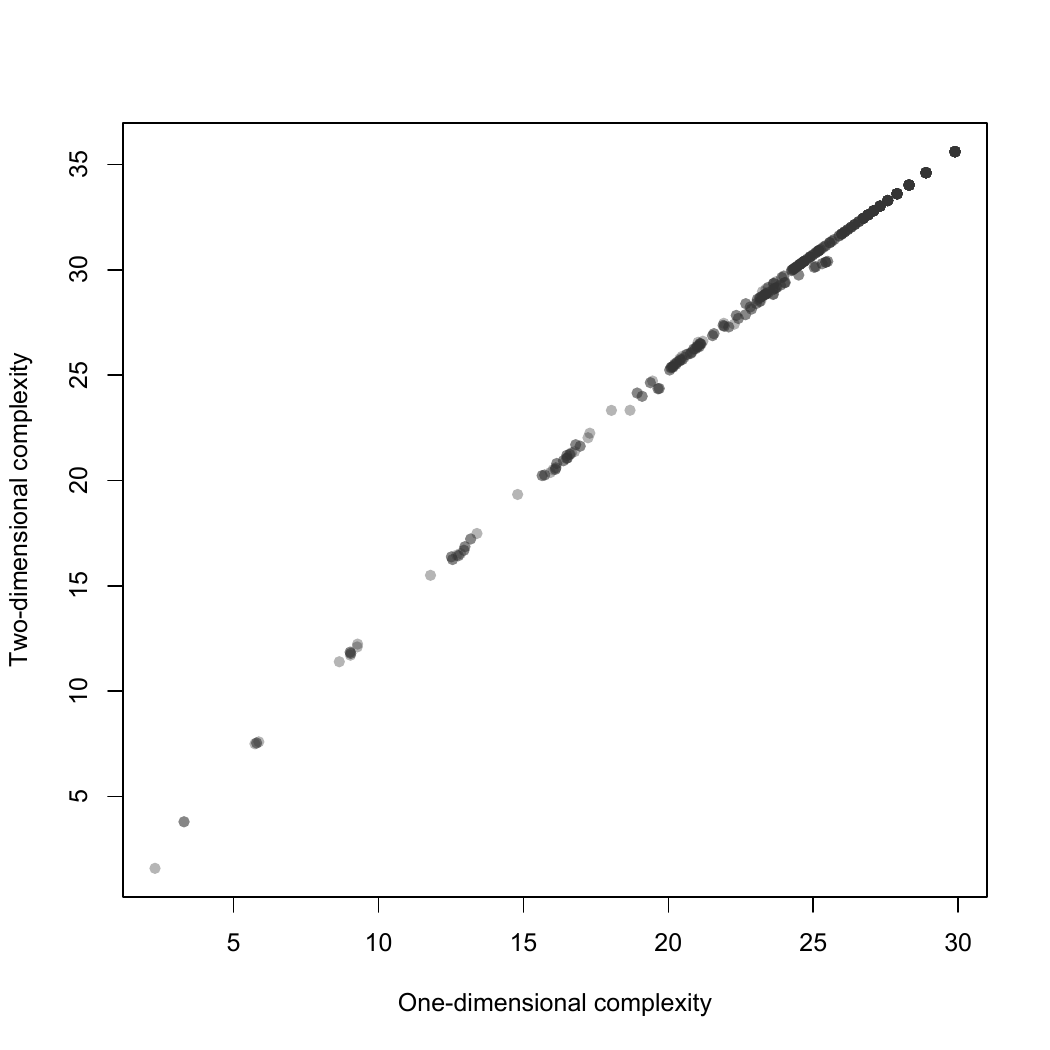}
\caption{Scatterplot of $K_{m}$ with 2-dimensional Turing machines (Turmites) as a function of $K_{m}$ with 1-dimensional Turing machines.}
\label{BasicScatterplot}
\end{center}
\end{figure}

We shall now look at how 1-dimensional arrays (hence strings) produced by 2D Turing machines correlate with strings that we have calculated before \cite{thesis,delahayezenil,d5} (denoted by $D(5)$). In a sense this is like changing the Turing machine formalism to see whether the new distribution resembles distributions following other Turing machine formalisms, and whether it is robust enough. 

All Turing machines in $(4,2)$ are included in $(4,2)_{2D}$ because these are just the machines that do not move up or down. We first compared the values of the 1832 output strings in $(4,2)$ to the 1-dimensional arrays found in $(4,2)_{2D}$. We are also interested in the relation between the ranks of these 1832 strings in both $(4,2)$ and $(4,2)_{2D}$. 

\begin{figure}[htbp]
\begin{center}
 \includegraphics[width=13.5cm]{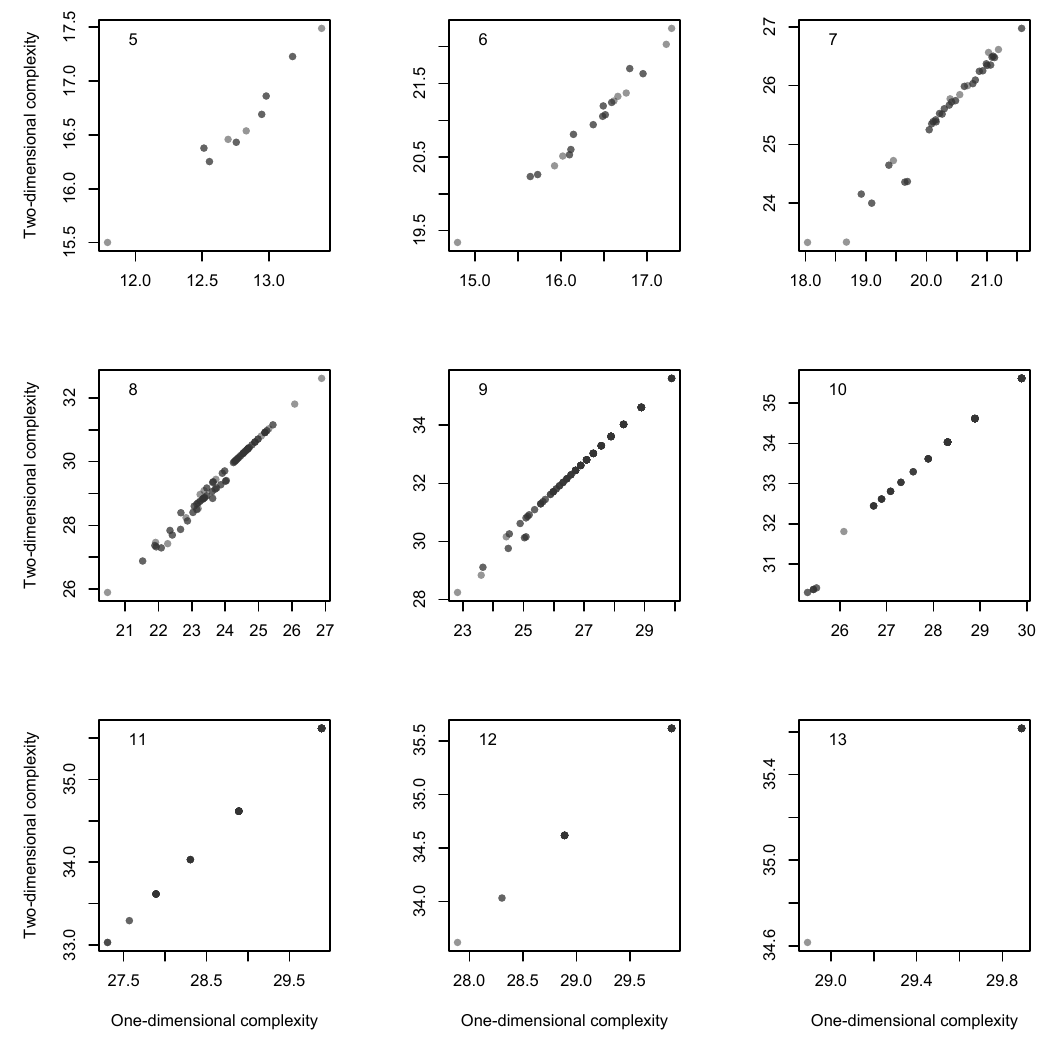}
\caption{Scatterplot of $K_{m}$ with 2-dimensional Turing machines as a function of $K_{m}$ with 1-dimensional Turing machines by length of strings, for strings of length 5 to 13.} 
\label{ScatterplotByLength}
\end{center}
\end{figure}

Fig.~\ref{BasicScatterplot} shows the link between $K_{m,2D}$ with 2D Turing machines as a function of ordinary $K_{m,1D}$ (that is, simply $K_m$ as defined in \cite{d5}). It suggests a strong almost-linear overall association. The correlation coefficient $r=0.9982$ confirms the linear association, and the Spearman correlation coefficient $r_{s}=0.9998$ proves a tight and increasing functional relation.

The length $l$ of strings is a possible confounding factor. However Fig.~\ref{ScatterplotByLength} suggests that the link between one and 2-dimensional complexities is not explainable by $l$. Indeed, the partial correlation $r_{K_{m,1D}K_{m,2D}.l}=0.9936$ still denotes a tight association. 

Fig.~\ref{ScatterplotByLength} also suggests that complexities are more strongly linked with longer strings. This is in fact the case, as Table \ref{CorByLength} shows: the strength of the link increases with the length of the resulting strings. One and 2-dimensional complexities are remarkably correlated and may be considered two measures of the 
same underlying feature of the strings. How these measures vary is another matter. The regression of $K_{m,2D}$ on $K_{m,1D}$ gives the following approximate relation: $K_{m,2D} \approx 2.64+1.11K_{m,1D}.$ Note that this subtle departure from identity may be a consequence of a slight non-linearity, a feature visible in Fig.~\ref{BasicScatterplot}. 

\begin{table}[htdp]
\caption{Correlation coefficients between one and 2-dimensional complexities by length of strings.} 
\begin{center}
\begin{tabular}{cc}
\hline
Length ($l$) & Correlation\\
\hline
5 & 0.9724 \\
6 & 0.9863 \\
7 & 0.9845 \\
8 & 0.9944\\
9 & 0.9977\\
10 & 0.9952\\
11 & 1\\
12 & 1\\
\hline
\end{tabular}
\end{center}
\label{CorByLength}
\end{table}%

\subsection{Comparison of $K_m$ and compression of Cellular Automata}
\label{eca}

A 1-dimensional CA can be represented by an array of {\it cells} $x_i$ where $i \in \mathbb{Z}$ (integer set) and each $x$ takes a value from a finite alphabet $\Sigma$. Thus, a sequence of cells \{$x_i$\} of finite length $n$ describes a string or {\it global configuration} $c$ on $\Sigma$. This way, the set of finite configurations will be expressed as $\Sigma^n$. An evolution comprises a sequence of configurations $\{c_i\}$ produced by the mapping $\Phi:\Sigma^n \rightarrow \Sigma^n$; thus the global relation is symbolized as:

\begin{equation}
\Phi(c^t) \rightarrow c^{t+1}
\label{globalFunction}
\end{equation}

\noindent Where $t$ represents time and every global state of $c$ is defined by a sequence of cell states. The global relation is determined over the cell states in configuration $c^t$ updated simultaneously at the next configuration $c^{t+1}$ by a local function $\varphi$ as follows:

\begin{equation}
\label{ecafunction}
\varphi(x_{i-r}^t, \ldots, x_{i}^t, \ldots, x_{i+r}^t) \rightarrow x_i^{t+1}.
\end{equation}

Wolfram \cite{wolfram} represents 1-dimensional cellular automata (CA) with two parameters $(k,r)$ where $k = |\Sigma|$ is the number of states, and $r$ is the neighborhood radius. Hence this type of CA is defined by the parameters $(2,1)$. There are $\Sigma^n$ different neighborhoods (where $n=2r+1$) and $k^{k^n}$ distinct evolution rules. The evolutions of these cellular automata usually have periodic boundary conditions. Wolfram calls this type of CA Elementary Cellular Automata (denoted simply by ECA) and there are exactly $k^{k^n}=256$ rules of this type. They are considered the most simple cellular automata (and among the simplest computing programs) capable of great behavioral richness.

1-dimensional ECA can be visualized in 2-dimensional space-time diagrams where every row is an evolution in time of the ECA rule. By their simplicity and because we have a good understanding about them (e.g. at least one ECA is known to be capable of Turing universality \cite{cook,wolfram}) they are excellent candidates to test our measure $K_{m,2D}$, being just as effective as other methods that approach ECA using compression algorithms \cite{zenilca} that have yielded the results that Wolfram obtained heuristically.

\subsection{$K_{m,2D}$ comparison with compressed ECA evolutions}

We have seen that our Coding theorem method with associated measure $K_m$ (or $K_{m,2D}$ in this paper for 2D Kolmogorov complexity) is in agreement with bit string complexity as approached by compressibility, as we have reported in Section~\ref{sec:strings-lenghts-10}. 

The Universal Distribution from Turing machines that we have calculated ($D(4,2)_{2D}$) will help us to classify Elementary Cellular Automata. Classification of ECA by compressibility has been done before in \cite{zenilca} with results that are in complete agreement with our intuition and knowledge of the complexity of certain ECA rules (and related to Wolfram's classification \cite{wolfram}). In \cite{zenilca} both classifications by simplest initial condition and random initial condition were undertaken, leading to a stable compressibility classification of ECAs. Here we followed the same procedure for both simplest initial condition (single black cell) and random initial condition in order to compare the classification to the one that can be approximated by using $D(4,2)_{2D}$, as follows. 

We will say that the space-time diagram (or evolution) of an
Elementary Cellular Automaton $c$ after time $t$ has complexity: 

\begin{equation}
\label{eqeca}
K_{m,2D_{d\times d}}(c^t) = \sum_{q\in \{c^t\}_{d\times
  d}} K_{m,2D}(q)
\end{equation}

\noindent That is, the complexity of a cellular automaton $c$ is the
sum of the complexities of the $q$ arrays or image patches
in the partition matrix $\{c^t\}_{d\times d}$ from breaking $\{c^t\}$ into
square arrays of length $d$ produced by the ECA after $t$ steps. An
example of a partition matrix of an ECA evolution is shown in
Fig.~\ref{sample} for ECA Rule 30 and $d=3$ where
$t=6$. Notice that the boundary conditions for a partition matrix
may require the addition of at most $d-1$ empty rows or $d-1$ empty
columns to the boundary as shown in Fig.~\ref{sample} (or
alternatively the dismissal of at most $d-1$ rows or $d-1$ columns) if
the dimensions (height and width) are not multiples of $d$, in this
case $d=3$. 

\begin{figure}[!htb]
\begin{center}
\includegraphics[width=10.5cm]{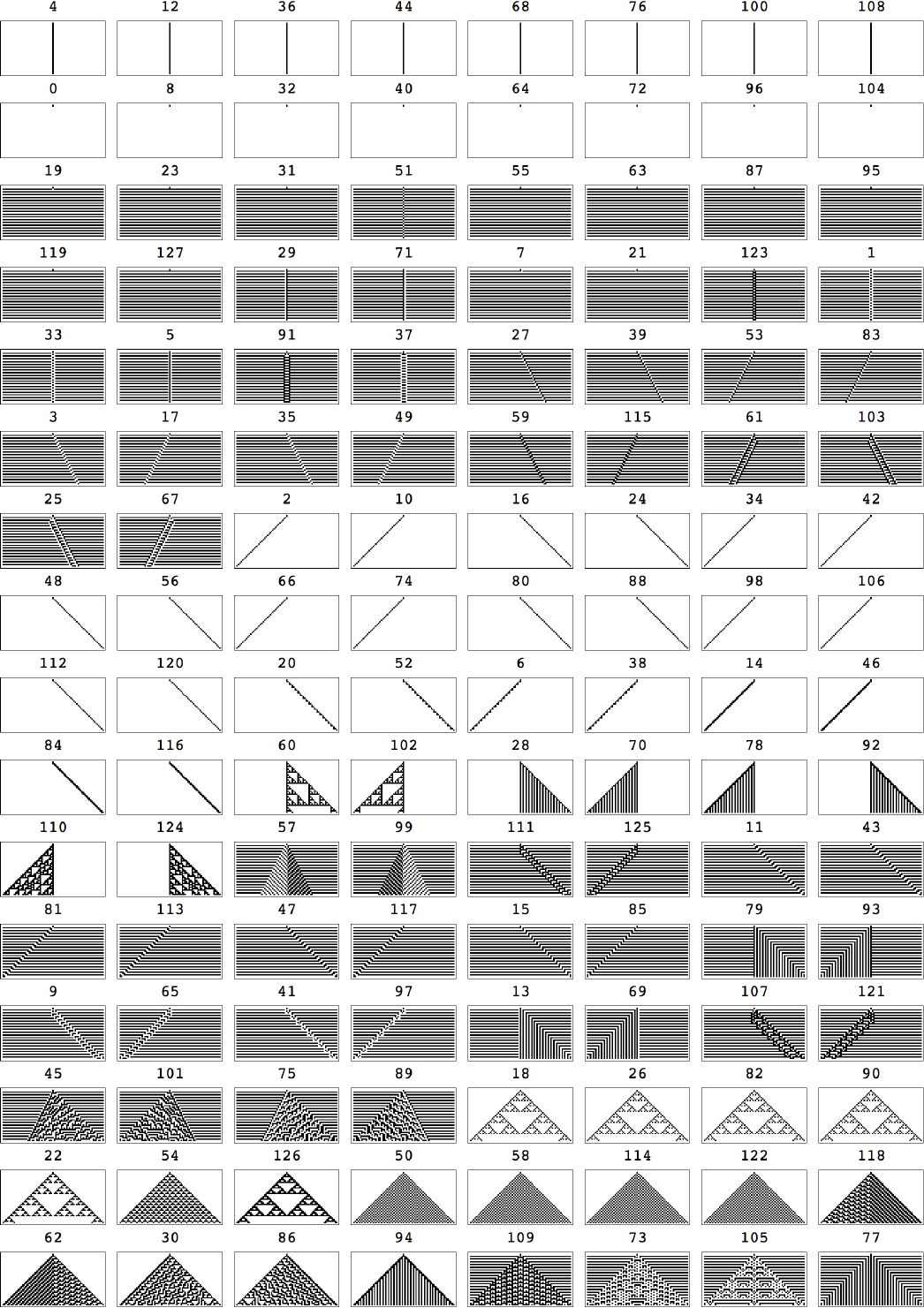}
\caption{\label{allecas} All the first 128 ECAs (the other 128 are 0-1 reverted rules) starting from the simplest (black cell) initial configuration running for $t=36$ steps, sorted from lowest to highest complexity according to $K_{m,2D_{3\times3}}$. Notice that the same procedure can be extended for its use on arbitrary images.}
\end{center}
\end{figure}

If the classification of all rules in ECA by $K_{m,2D}$ yields the same classification obtained by compressibility, one would be persuaded that $K_{m,2D}$ is a good alternative to compressibility as a method for approximating the Kolmogorov complexity of objects, with the signal advantage that $K_{m,2D}$ can be applied to very short strings and very short arrays such as images. Because all possible $2^9$ arrays of size $3\times 3$ are present in $K_{m,2D}$ we can use this arrays set to try to classify all ECAs by Kolmogorov complexity using the Coding Theorem method. Fig~\ref{fig:arrays3x3} shows all relevant (non-symmetric) arrays. We denote by $K_{m,2D_{3\times3}}$ this subset from $K_{m,2D}$.

\begin{figure}
\centering
\includegraphics[width=8cm]{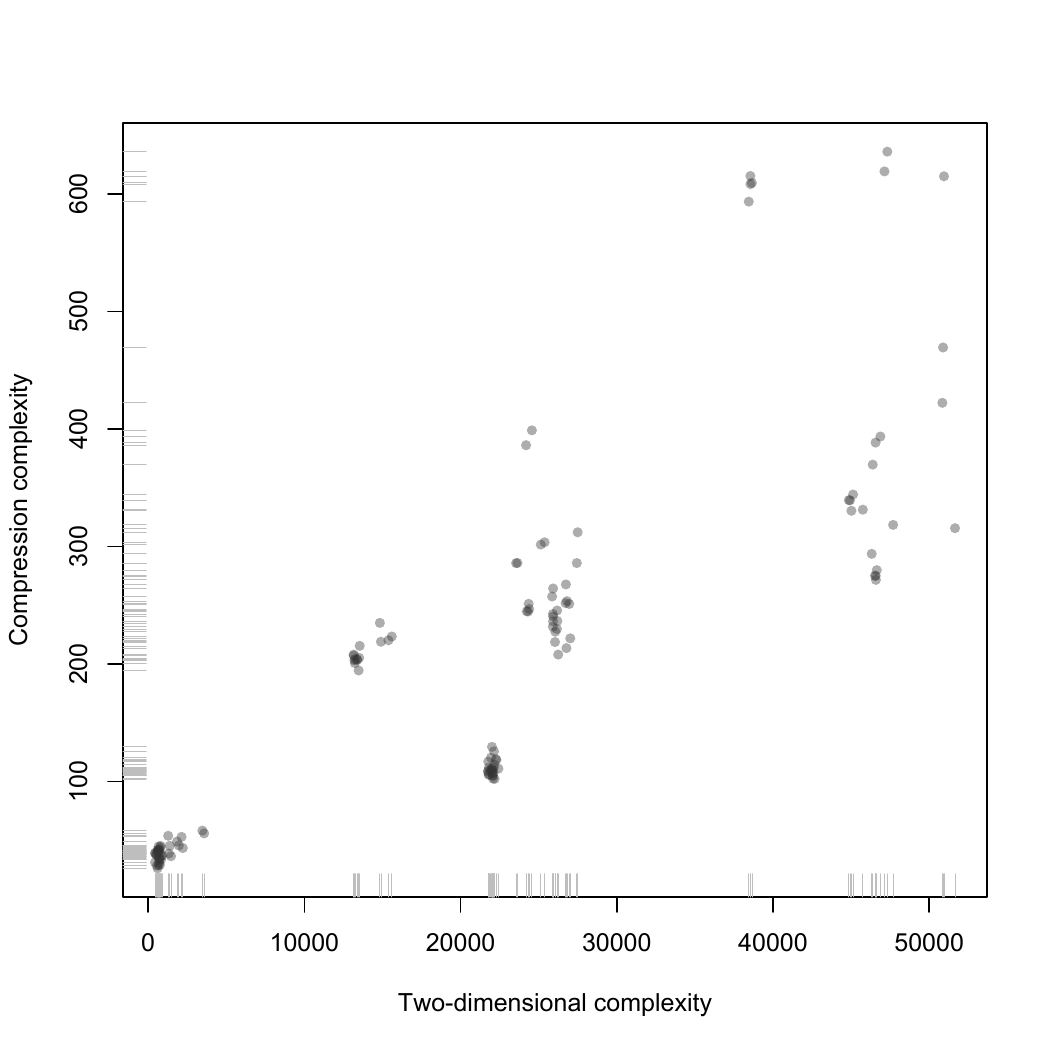}\\
\vspace{-1.2cm}
\includegraphics[width=8cm]{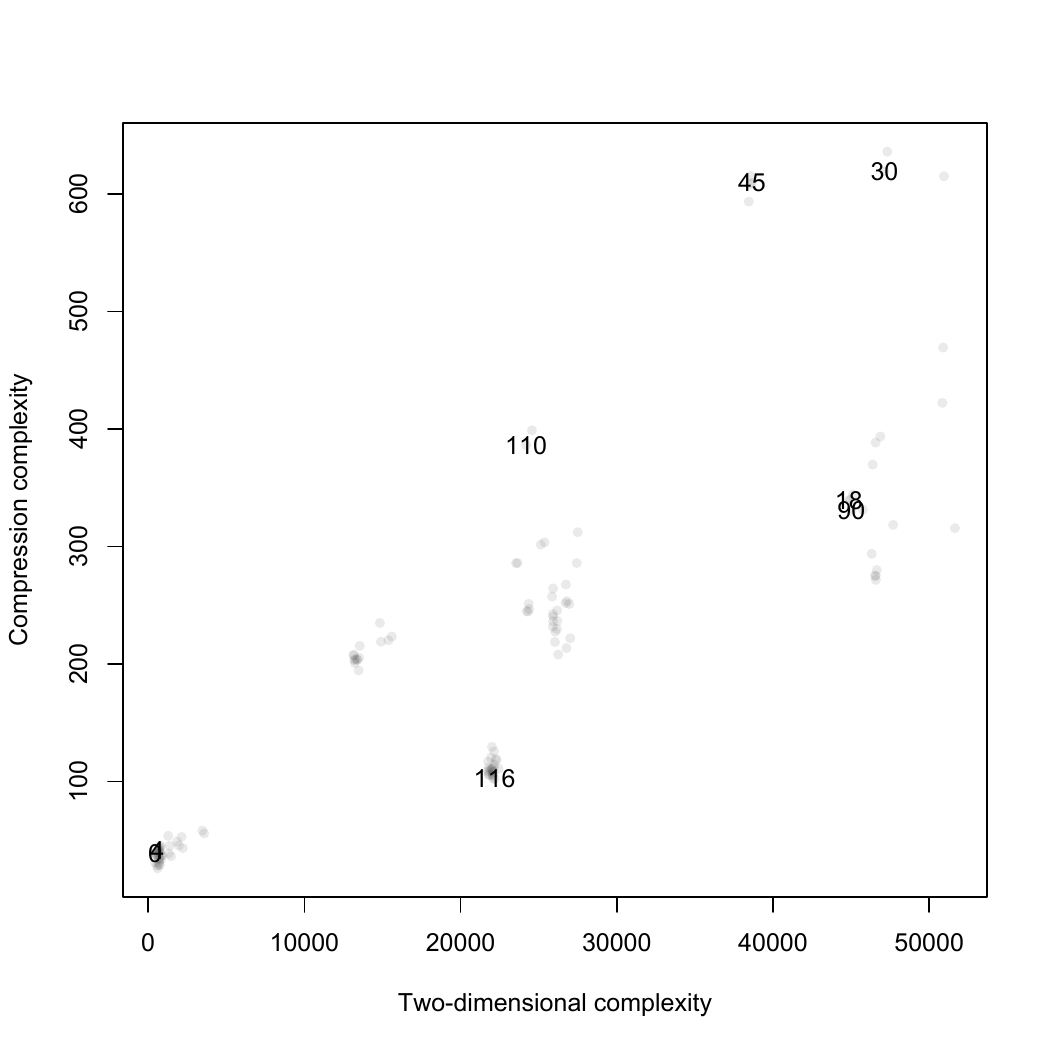}
\caption{\label{scatterplots} Scatterplots of Compress versus $K_{m,2D_{3\times3}}$ on the 128 first ECA evolutions after $t=90$ steps. Top: Distribution of points along the axes displaying clusters of equivalent rules and a distribution corresponding to the known complexity of various cases. Bottom: Same plot but with some ECA rules highlighted some of which were used in the side by side comparison in Fig.\ref{sample} (but unlike there, here for a single black cell initial condition). That rules distribute on the diagonal indicates that both methods are correlated as theoretically expected (even if lossless compression is a form of entropy rate up to the compression fixed maximum word length).}
\end{figure}

Fig.~\ref{scatterplots} displays the scatterplot of compression complexity against $K_{m,2D_{3\times3}}$ calculated for every cellular automaton. It shows a positive link between the two measures. The Pearson correlation amounts to $r=0.8278$, so the determination coefficient is $r^{2}=0.6853$. These values correspond to a strong correlation, although smaller than the correlation between 1- and 2-dimensional complexities calculated in Section~\ref{sec:strings-lenghts-10}.

Concerning orders arising from these measures of complexity, they too are strongly linked, with a Spearman correlation of $r_{s}=0.9200$. The scatterplots (Fig.~\ref{scatterplots}) show a strong agreement between the Coding theorem method and the traditional compression method when both are used to classify ECAs by their approximation to Kolmogorov complexity.

The anomalies found in the classification of Elementary Cellular Automata (e.g. Rule 77 being placed among ECA with high complexity according to $K_{m,2D_{3\times3}}$) is a limitation of $K_{m,2D_{3\times3}}$ itself and not of the Coding theorem method which for $d=3$ is unable to ``see" beyond 3-bit squares using, which is obviously very limited. And yet the degree of agreement with compressibility is surprising (as well as with intuition, as a glance at Fig.~\ref{allecas} shows, and as the distribution of ECAs starting from random initial conditions in Fig.~\ref{sample} confirms). In fact an average ECA has a complexity of about 20K bits, which is quite a large program-size when compared to what we intuitively gauge to be the complexity of each ECA, which may suggest that they should have smaller programs. However, one can think of $D(4, 2)_{2D_{3 \times 3}}$ as attempting to reconstruct the evolution of each ECA for the given number of steps with square arrays only 3 bits in size, the complexity of the three square arrays adding up to approximate $K_{m,2D}$ of the ECA rule. Hence it is the deployment of $D(4, 2)_{2D_{3 \times 3}}$ that takes between 500 to 50K bits to reconstruct every ECA space-time evolution depending on how random vs. how simple it is. 

Other ways to exploit the data from $D(4, 2)_{2D}$ (e.g. non-square arrays) can be utilized to explore better classifications. We think that constructing a Universal Distribution from a larger set of Turing machines, e.g. $D(5, 2)_{2D_{4 \times 4}}$ will deliver more accurate results but here we will also introduce a tweak to the definition of the complexity of the evolution of a cellular automaton.

\begin{figure}[!htb]
\begin{center}
\includegraphics[width=10.9cm]{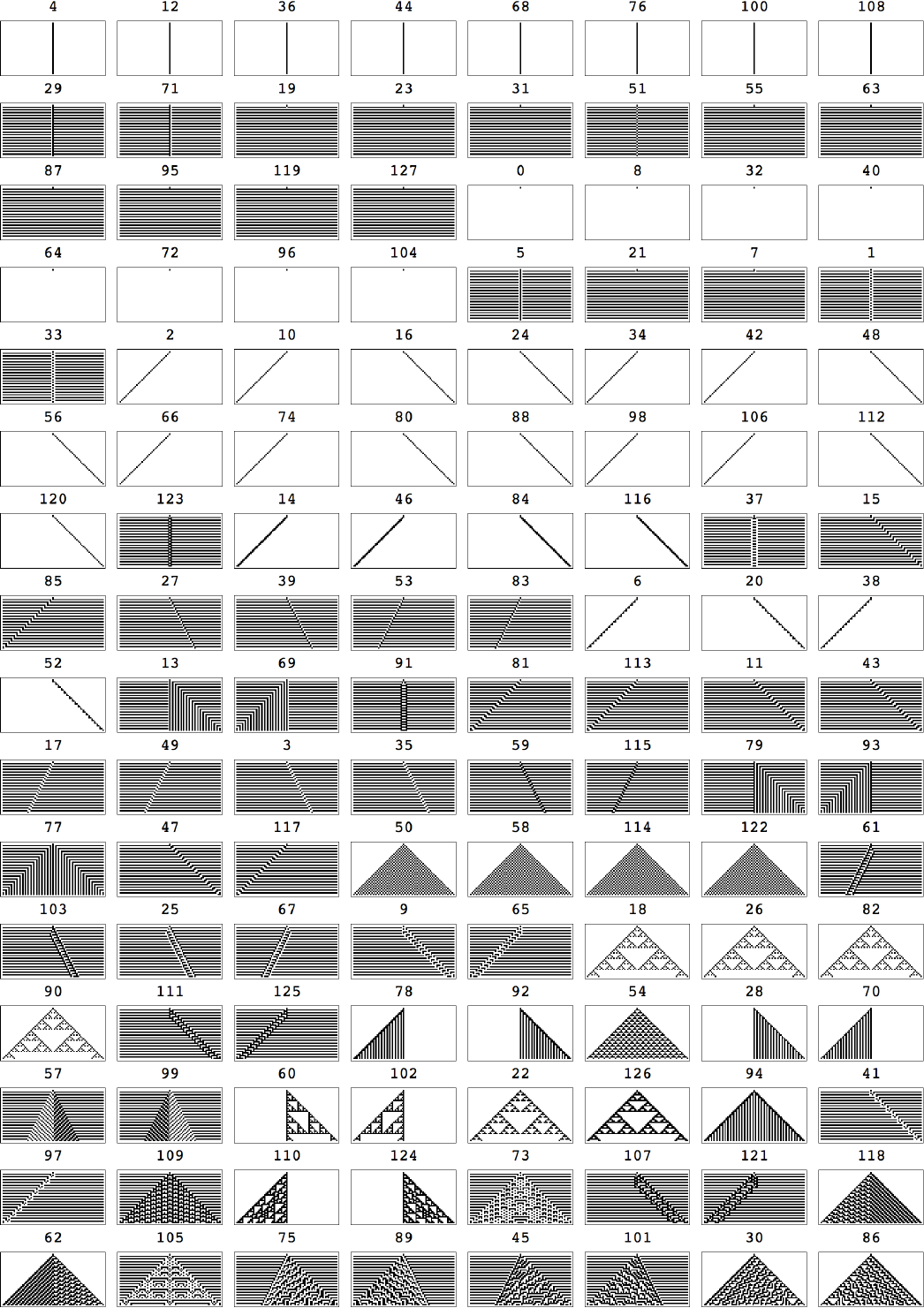}
\caption{\label{allecaslog} \emph{Block Decomposition Method}. All the first 128 ECAs (the other 128 are 0-1 reverted rules) starting from the simplest (black cell) initial configuration running for $t=36$ steps, sorted from lowest to highest complexity according to $Klog$ as defined in Eq.~\ref{newecaeq}.}
\end{center}
\end{figure}

Splitting ECA rules in array squares of size 3 is like trying to look through little windows 9 pixels wide one at a time in order to recognize a face, or training a ``microscope'' on a planet in the sky. One can do better with the Coding theorem method by going further than we have in the calculation of a 2-dimensional Universal Distribution (e.g. calculating in full or a sample of $D(5, 2)_{2D_{4 \times 4}}$), but eventually how far this process can be taken is dictated by the computational resources at hand. Nevertheless, one should use a telescope where telescopes are needed and a microscope where microscopes are needed.

\subsection{Block Decomposition Method}

One can think of an improvement in resolution of $K_{m,2D}(c)$ for growing space-time diagrams of cellular automaton by taking the $\log_2(n)$ of the sum of the arrays where $n$ is the number of repeated arrays, instead of simply adding the complexity of the image patches or arrays. That is, one penalizes repetition to improve the resolution of $K_{m,2D}$ for larger images as a sort of ``optical lens". This is possible because we know that the Kolmogorov complexity of repeated objects grows by $\log_2(n)$, just as we explained with an example in Section~\ref{kolmo}. Adding the complexity approximation of each array in the partition matrix of a space-time diagram of an ECA provides an upper bound on the ECA Kolmogorov complexity, as it shows that there is a program that generates the ECA evolution picture with the length equal to the sum of the programs generating all the sub-arrays (plus a small value corresponding to the code length to join the sub-arrays). So if a sub-array occurs $n$ times we do not need to consider it's complexity $n$ times but $\log_2(n)$. Taking into account this, Eq.~\ref{eqeca} can be then rewritten as:

\begin{equation}
\label{newecaeq}
K^\prime_{m,2D_{d\times d}}(c^t) = \sum_{(r_u,n_u)\in \{c^t\}_{d\times d}} K_m(r_u) + \log_2(n_u)
%Klog_{m,2D_{d\times d}}(c^t) = \sum_{r_u\in \{c^t\}_{d\times d}} \log_2(n) K_{m,2D}(r_u) + K_{m,2D}(r_u)
\end{equation}

\noindent Where $r_u$ are the different square arrays in the
  partition $\{c^t\}_{d\times d}$ of the matrix $c^t$ and $n_u$ the
  multiplicity of $r_u$, that is the number of repetitions of $d\times
  d$-length patches or square arrays found in $c^t$. From now on
  we will use $K^\prime$ for squares of size greater than 3 and it may
  be denoted only by $K$ or by BDM standing for \emph{Block decomposition
    method}. BDM has now been applied successfully to measure, for example, the
  Kolmogorov complexity of graphs and complex networks~\cite{zenilphysicaA} by way of their adjacency matrices (a 2D grid) and was shown to be consistent with labelled and unlabelled (up to isomorphisms) graphs.

\begin{figure}[!htb]
\begin{center}
\includegraphics[width=5.9cm]{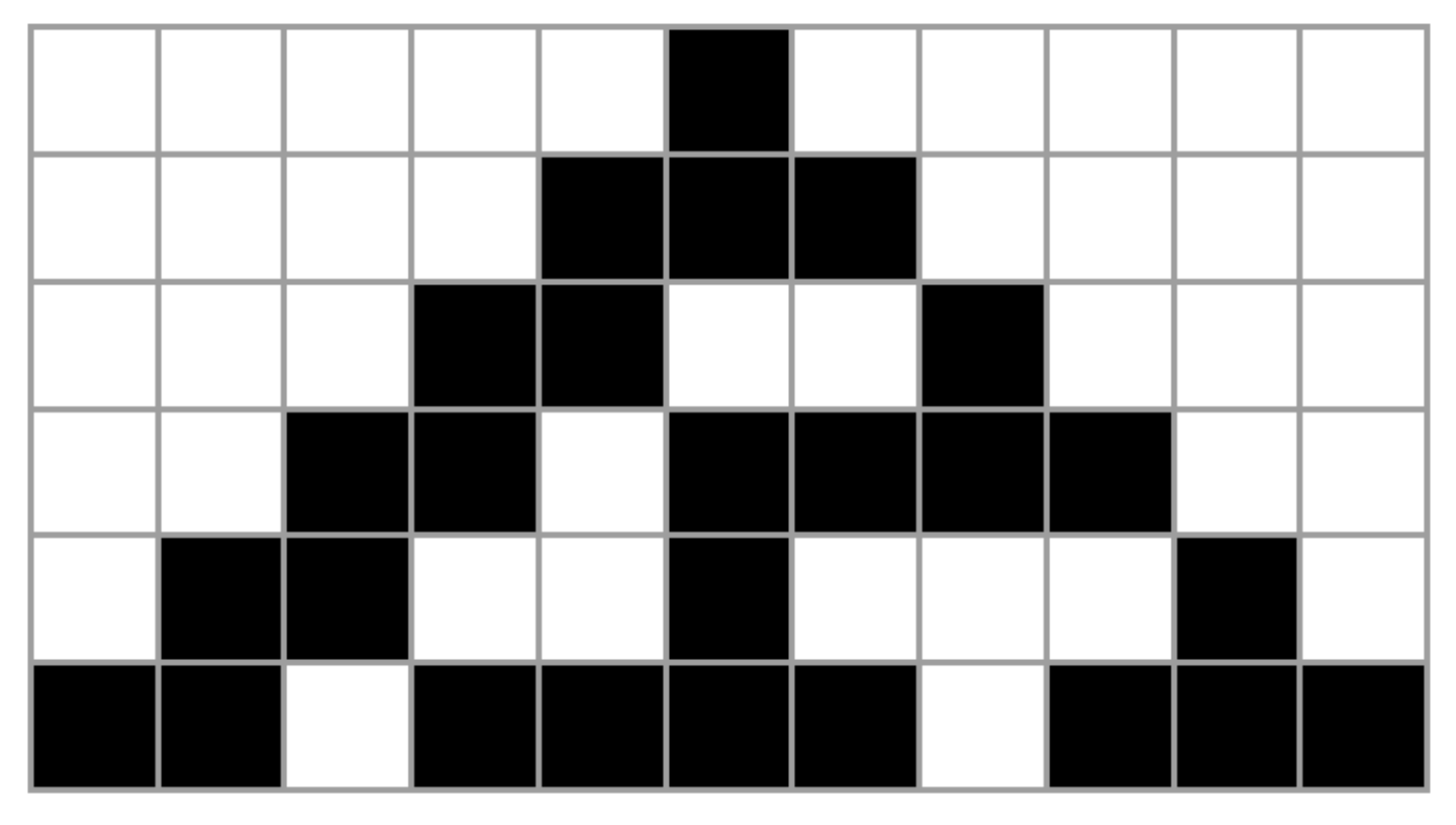}\bigskip

$\Downarrow$\bigskip

\includegraphics[width=8.2cm]{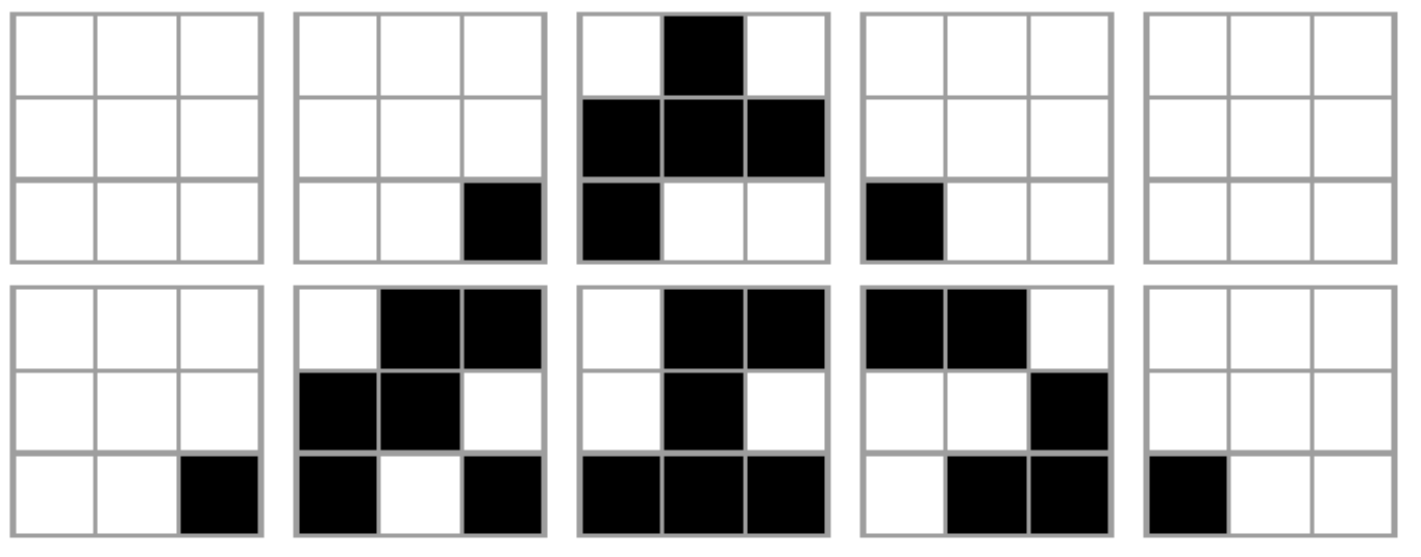}\\
\vspace{.3cm}

\noindent\rule{6cm}{0.4pt}\\

\vspace{.5cm}
\includegraphics[width=13cm]{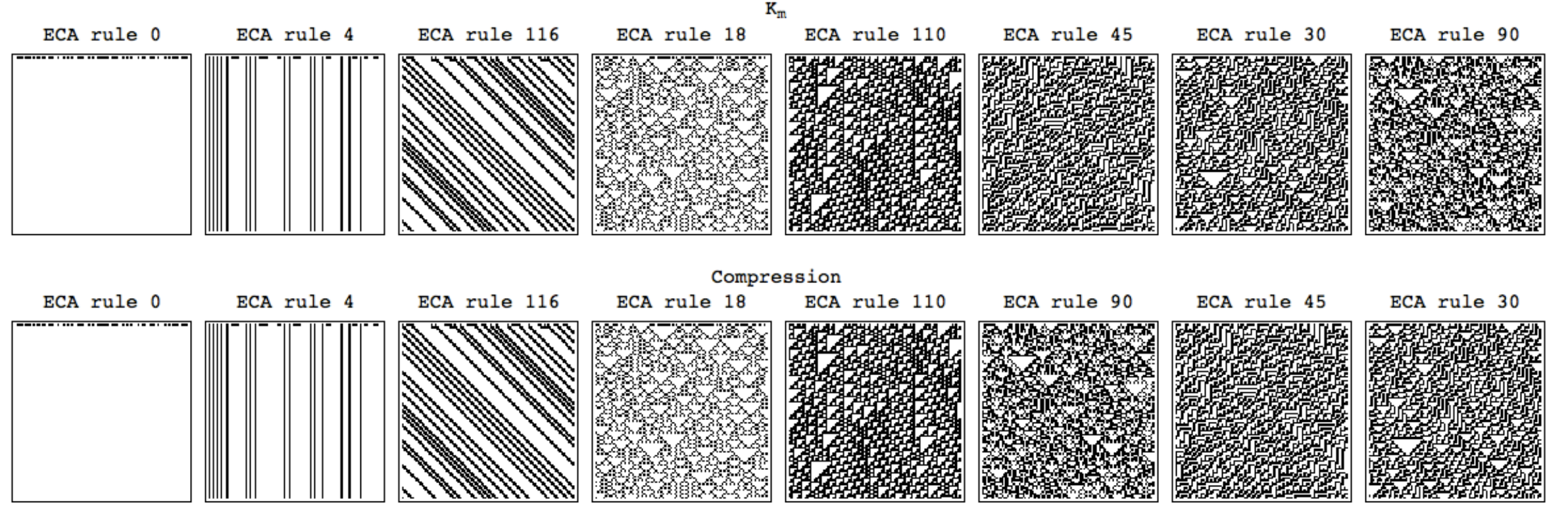}
\caption{\label{sample} Top: Block decomposing (other boundary conditions are possible and under investigation) the evolution of Rule 30 (top) ECA after $t=6$ steps into 10 subarrays of length $3\times3$ (bottom) in order to calculate $K_{m,2D_{3\times3}}$ to approximate its Kolmogorov complexity. Bottom: Side by side comparison of 8 evolutions of representative ECAs, starting from a random initial configuration, sorted from lowest to highest BDM values (top) and smallest to largest compression lengths using the Deflate algorithm as a method to approximate Kolmogorov complexity \cite{zenilca}.}
\end{center}
\end{figure}

Now complexity values of $K^\prime_{m,2D_{d\times d}}$ range between 70 to 3K bits with a mean program-size value of about 1K bits. The classification of ECA, according to Eq.~\ref{newecaeq}, is presented in Fig.~\ref{allecaslog}. There is an almost perfect agreement with a classification by lossless compression length (see Fig.~\ref{sample}) which makes even one wonder whether the Coding theorem method is actually providing more accurate approximations to Kolmogorov complexity than lossless compressibility for this objects length. Notice that the same procedure can be extended for its use on arbitrary images. We denominate this technique \emph{Block Decomposition Method}. We think it will prove to be useful in various areas, including machine learning as an of Kolmogorov complexity (other contributions to ML inspired in Kolmogorov complexity can be found in \cite{hutter}).

Also worth notice that the fact that ECA can be successfully classified by $K_{m,2D}$ with an approximation of the Universal Distribution calculated from Turing machines (TM) suggests that output frequency distributions of ECA and TM cannot be but strongly correlated, something that we had found and reported before in \cite{zenilalgo} and \cite{zenil2007}.

Another variation of the same $K_{m,2D}$ measure is to divide the original image into all possible square arrays of a given length rather than taking a partition. This would, however, be exponentially more expensive than the partition process alone, and given the results in Fig.~\ref{allecaslog} further variations do not seem to be needed, at least not for this case.

\subsection{Robustness of the approximations to $m(s)$}

One important question that arises when positing the soundness of the Coding theorem method as an alternative to having to pick a universal Turing machine to evaluate the Kolmogorov complexity $K$ of an object, is how many arbitrary choices are made in the process of following one or another method and how important they are. One of the motivations of the Coding theorem method is to deal with the constant involved in the Invariance theorem (Eq.~\ref{invariance}), which depends on the (prefix-free) universal Turing machine chosen to measure $K$ and which has such an impact on real-world applications involving short strings. While the constant involved remains, given that after application of the Coding theorem (Eq.~\ref{coding}) we reintroduce the constant in the calculation of $K$, a legitimate question to ask is what difference it makes to follow the Coding theorem method compared to simply picking the universal Turing machine.

On the one hand, one has to bear in mind that no other method existed for approximating the Kolmogorov complexity of short strings. On the other hand, we have tried to minimize any arbitrary choice, from the formalism of the computing model to the informed runtime, when no Busy Beaver values are known and therefore sampling the space using an educated runtime cut-off is called for. When no Busy Beaver values are known the chosen runtime is determined according to the number of machines that we are ready to miss (e.g. less than .01\%) for our sample to be significative enough as described in Section~\ref{sec:setting-runtime}. We have also shown in~\cite{d5} that approximations to the Universal Distribution from spaces for which Busy Beaver values are known are in agreement with larger spaces for which Busy Beaver values are not known.

Among the possible arbitrary choices it is the enumeration that may perhaps be questioned, that is, calculating $D(n)$ for increasing $n$ (number of Turing machine states), hence by increasing size of computer programs (Turing machines). On the one hand, one way to avoid having to make a decision on the machines to consider when calculating a Universal Distribution is to cover all of them for a given number of $n$ states and $m$ symbols, which is what we have done (hence the enumeration in a thoroughly $(n, m)$ space becomes irrelevant). While it may be an arbitrary choice to fix $n$ and $m$, the formalisms we have followed guarantee that  $n$-state $m$-symbol Turing machines are in $(n+i, m+j)$ with $i, j \geq 0$ (that is, the space of all $n+i$-state $m+j$-symbol Turing machines). Hence the process is incremental, taking larger spaces and constructing an average Universal Distribution. In fact, we have demonstrated \cite{d5} that $D(5)$ (that is, the Universal Distribution produced by the Turing machines with 2 symbols and 5 states) is strongly correlated to $D(4)$ and represents an improvement in accuracy of the string complexity values in $D(4)$, which in turn is in agreement with and an improvement on $D(3)$ and so on. We have also estimated the constant $c$ involved in the invariance theorem (Eq.~\ref{invariance}) between these $D(n)$ for $n>2$, which turned out to be very small in comparison to all the other calculated Universal Distributions \cite{numerical}. 

%The invariance theorem guarantees that this $c$ will eventually tend to zero for $n\rightarrow \infty$. However, it doesn't indicate the rate of the convergence, but the fact that $c$ remains very small among the calculated Universal Distributions is a good sign.

\subsection{Real-world evidence}

We have provided here some theoretical and statistical arguments to show the reliability, validity and generality of our measure, more empirical evidence has also been produced, in particular in the field of cognition and psychology where researchers often have to deal with too short strings or too small patterns for compression methods to be used. For instance, it was found that the complexity of a (one-dimensional) string better predicts its recall from short-term memory that the length of the string~\cite{chekaf15}. Incidentally, a study on the
conspiracy theory believers mindset also revealed that human perception of randomness is highly linked to our one-dimensional measure of complexity~\cite{dieguez15}. Concerning the two-dimensional version introduced in this paper, it has been fruitfully used to show how language iterative learning triggers the emergence of linguistic structures~\cite{kempe15}. A direct link between the perception of two-dimensional randomness, our complexity measure, and natural statistics was also established in two experiments~\cite{gauvrit14}. These findings further support the complexity metrics presented herein. Furthermore, more theoretical arguments have been advanced in~\cite{numerical} and ~\cite{finite}.

\section{Conclusions}

%We have shown that the results of $K_{m,2D}$ are in agreement with compressibility when applied to a concatenation of strings of the same complexity, and to classifications of Elementary Cellular Automata (ECA) by compressibility. However, $K_{m,2D}$ (like our $K_m$ in \cite{d5}) is finer-grained compared to the number of different classes provided by compressibility. As a consequence the Coding theorem method seems to have the advantage over lossless compression algorithms to better distinguish complexity of small objects. For example, among the ECA, $K_{m,2D_{3 \times 3}}$ provided 51 different complexity values, while compression retrieved 45 different values. These are not very different from each other but the smaller the object the more important it is to have a greater number of complexity values in order to distinguish the complexity of one string from another. For short enough objects compression algorithms collapse all complexity values into a random value (the compressed length of short objects is larger than the original object itself) as seen in \cite{d5}. Together with the evidence we have provided in this paper, our measure and the Coding theorem method we are advancing here look sound and ready for applications, ready to serve as a complement to the compression method, especially where compression fails. 

We have shown how a highly symmetric but algorithmic process is capable of generating a full range of patterns of different structural complexity. We have introduced this technique as a natural and objective measure of complexity for $n$-dimensional objects. With two different experiments we have demonstrated that the measure is compatible with lossless compression estimations of Kolmogorov complexity, yielding similar results but providing an alternative particularly for short strings. We have also shown that $K_{m,2D}$ (and $K_m$) are ready for applications, and that calculating Universal Distributions is a stable alternative to compression and a potential useful tool for approximating the Kolmogorov complexity of objects, strings and images (arrays). We think this method will prove to do the same for a wide range of areas where compression is not an option given the size of strings involved.

We also introduced the \emph{Block Decomposition Method}. As we have seen with anomalies in the classification such as ECA Rule 77 (see Fig.~\ref{allecas}), when approaching the complexity of the space-time diagrams of ECA by splitting them in square arrays of size 3, the Coding theorem method does have its limitations, especially because it is computationally very expensive (although the most expensive part needs to be done only once---that is, producing an approximation of the Universal Distribution). Like other high precision instruments for examining the tiniest objects in our world, measuring the smallest complexities is very expensive, just as the compression method can also be very expensive for large amounts of data.

We have shown that the method is stable in the face of the changes in Turing machine formalism that we have undertaken (in this case Turmites) as compared to, for example, traditional 1-dimensional Turing machines or to strict integer value program-size complexity~\cite{numerical} as a way to estimate the error of the numerical estimations of Kolmogorov complexity through algorithmic probability. For the Turing machine model we have now changed the number of states, the number of symbols and now even the movement of the head and its support (grid versus tape). We have shown and reported here and in~\cite{d5,numerical} that all these changes yield distributions that are strongly correlated with each other up to the point to assert that all these parameters have marginal impact in the final distributions suggesting a fast rate of convergence in values that reduce the concern of the constant involved in the invariance theorem. In~\cite{zenilalgo} we also proposed a way to compare approximations to the Universal Distribution by completely different computational models (e.g. Post tag systems and cellular automata), showing that for the studied cases reasonable estimations with different degrees of correlations were produced. The fact that we classify Elementary Cellular Automata (ECA) as shown in this paper, with the output distribution of Turmites with results that fully agree with lossless compressibility, can be seen as evidence of agreement in the face of a radical change of computational model that preserves the apparent order and randomness of Turmites in ECA and of ECA in Turmites, which in turn are in full agreement with 1-dimensional Turing machines and with lossless compressibility.

We have made available to the community this ``microscope" to look at the space of bit strings and other objects in the form of the Online Algorithmic Complexity Calculator (\url{http://www.complexitycalculator.com}) implementing $K_m$ (in the future it will also implement $K_{m,2D}$ and many other objects and a wider range of methods) that provides objective algorithmic probability and Kolmogorov complexity estimations for short binary strings using the method described herein. Raw data and the computer programs to reproduce the results for this paper can also be found under the Publications section of the Algorithmic Nature Group (\url{http://www.algorithmicnature.org}).

\end{document}